\documentclass[]{revtex4}
\usepackage{graphicx}
\usepackage{fancyhdr}

\begin{document}

\def\ds{\displaystyle}
\def\beq{\begin{equation}}
\def\eeq{\end{equation}}
\def\bea{\begin{eqnarray}}
\def\eea{\end{eqnarray}}
\def\beeq{\begin{eqnarray}}
\def\eeeq{\end{eqnarray}}
\def\ve{\vert}
\def\vel{\left|}
\def\ver{\right|}
\def\nnb{\nonumber}
\def\ga{\left(}
\def\dr{\right)}
\def\aga{\left\{}
\def\adr{\right\}}
\def\lla{\left<}
\def\rra{\right>}
\def\rar{\rightarrow}
\def\nnb{\nonumber}
\def\la{\langle}
\def\ra{\rangle}
\def\ba{\begin{array}}
\def\ea{\end{array}}
\def\tr{\mbox{Tr}}
\def\ssp{{\Sigma^{*+}}}
\def\sso{{\Sigma^{*0}}}
\def\ssm{{\Sigma^{*-}}}
\def\xis0{{\Xi^{*0}}}
\def\xism{{\Xi^{*-}}}
\def\qs{\la \bar s s \ra}
\def\qu{\la \bar u u \ra}
\def\qd{\la \bar d d \ra}
\def\qq{\la \bar q q \ra}
\def\gGgG{\la g^2 G^2 \ra}
\def\q{\gamma_5 \not\!q}
\def\x{\gamma_5 \not\!x}
\def\g5{\gamma_5}
\def\sb{S_Q^{cf}}
\def\sd{S_d^{be}}
\def\su{S_u^{ad}}
\def\rl{\hat{m}_{\ell}}
\def\ss{\hat{s}}
\def\rr{\hat{r}_{K_1}}
\def\sbp{{S}_Q^{'cf}}
\def\sdp{{S}_d^{'be}}
\def\sup{{S}_u^{'ad}}
\def\ssp{{S}_s^{'??}}
\def\sig{\sigma_{\mu \nu} \gamma_5 p^\mu q^\nu}
\def\fo{f_0(\frac{s_0}{M^2})}
\def\ffi{f_1(\frac{s_0}{M^2})}
\def\fii{f_2(\frac{s_0}{M^2})}
\def\O{{\cal O}}
\def\sl{{\Sigma^0 \Lambda}}
\def\es{\!\!\! &=& \!\!\!}
\def\ar{&+& \!\!\!}
\def\ek{&-& \!\!\!}
\def\cp{&\times& \!\!\!}
\def\fhs{Re[FH^*]}
\def\ghs{Re[GH^*]}
\def\bcs{Re[BC^*]}
\def\fgs{Re[FG^*]}
\def\hh{|H|^2}
\def\cc{|C|^2}
\def\gg{|G|^2}
\def\ff{|F|^2}
\def\bb{|B|^2}
\def\aa{|A|^2}
\def\hh{|H|^2}
\def\ee{|E|^2}
\def\rl{\hat{m}_{\ell}}
\def\r{\hat{m}_K}
\def\s{\hat{s}}
\def\ll{\Lambda}
\def\bcds{$B_c^\ast \rightarrow D_{s}~l^+ l^- $ }
\def\bcs{$B_c^\ast $}

\renewcommand{\textfraction}{0.2}    
\renewcommand{\topfraction}{0.8}

\renewcommand{\bottomfraction}{0.4}
\renewcommand{\floatpagefraction}{0.8}
\newcommand\mysection{\setcounter{equation}{0}\section}
\newcommand{\bra}[1]{\langle {#1}}
\newcommand{\ket}[1]{{#1} \rangle}
\newcommand{\ebar}{{\bar{e}}}
\newcommand{\sbar}{\bar{s}}
\newcommand{\cbar}{\bar{c}}
\newcommand{\bbar}{\bar{b}}
\newcommand{\qbar}{\bar{q}}
\renewcommand{\l}{\ell}
\newcommand{\lbar}{\bar{\ell}}
\newcommand{\psibar}{\bar{\psi}}
\newcommand{\barB}{\overline{B}}
\newcommand{\barK}{\overline{K}}
\newcommand{\thetaK}{\theta_{K_1}}
\newcommand{\onepone}{{1^1P_1}}
\newcommand{\sanpone}{{1^3P_1}}
\newcommand{\kone}{{K_1}}
\newcommand{\barkone}{{\overline{K}_1}}
\renewcommand{\Re}{\mathop{\mbox{Re}}}
\renewcommand{\Im}{\mathop{\mbox{Im}}}
\newcommand{\T}{{\cal T}}
\newcommand{\eff}{{\rm eff}}
\newcommand{\A}{{\cal A}}
\newcommand{\B}{{\cal B}}
\newcommand{\C}{{\cal C}}
\newcommand{\D}{{\cal D}}
\newcommand{\E}{{\cal E}}
\newcommand{\F}{{\cal F}}
\newcommand{\G}{{\cal G}}
\renewcommand{\H}{{\cal H}}
\newcommand{\hats}{\hat{s}}
\newcommand{\hatp}{\hat{p}}
\newcommand{\hatq}{\hat{q}}
\newcommand{\hatm}{\hat{m}}
\newcommand{\hatu}{\hat{u}}
\newcommand{\alphaem}{\alpha_{\rm em}}
\newcommand{\konel}{K_1(1270)}
\newcommand{\koneh}{K_1(1400)}
\newcommand{\barkonel}{\barK_1(1270)}
\newcommand{\barkoneh}{\barK_1(1400)}
\newcommand{\konea}{K_{1A}}
\newcommand{\koneb}{K_{1B}}
\newcommand{\barkonea}{\barK_{1A}}
\newcommand{\barkoneb}{\barK_{1B}}
\newcommand{\mkone}{m_{\kone}}
\newcommand{\konep}{K_1^+}
\newcommand{\konem}{K_1^-}
\newcommand{\konelm}{K_1^-(1270)}
\newcommand{\konehm}{K_1^-(1400)}
\newcommand{\konelp}{K_1^+(1270)}
\newcommand{\konehp}{K_1^+(1400)}
\newcommand{\konelz}{\overline{K}{}^0_1(1270)}
\newcommand{\konehz}{\overline{K}{}^0_1(1400)}
\newcommand{\Bm}{B^-}
\newcommand{\Bz}{\overline{B}{}^0}
\newcommand{\Kstar}{K^*(892)}
\newcommand{\BABAR}{BABAR}
\newcommand{\BELLE}{Belle}
\newcommand{\CLEO}{CLEO}
\newcommand{\leftu}{\gamma^\mu L}
\newcommand{\leftd}{\gamma_\mu L}
\newcommand{\rightu}{\gamma^\mu R}
\newcommand{\rightd}{\gamma_\mu R}
\newcommand{\Br}{{\cal B}}
\newcommand{\sect}[1]{Sec.~\ref{#1}}
\newcommand{\eqref}[1]{(\ref{#1})}
\newcommand{\fig}{FIG.~}
\newcommand{\figs}{FIGs.~}
\newcommand{\tbl}{TABLE~}
\newcommand{\tbls}{TABLEs~}
\newcommand{\errpm}[3]{#1^{+{#2}}_{-{#3}}}
\newcommand{\errpmf}[5]{{#1}^{ +{#2} +{#4} }_{-{#3}-{#5}}}
\newcommand{\lpm}{\l^+\l^-}
\newcommand{\epm}{e^+e^-}
\newcommand{\mupm}{\mu^+\mu^-}
\newcommand{\taupm}{\tau^+\tau^-}
\newcommand{\AFB}{A_{\rm FB}}
\newcommand{\barAFB}{\overline{A}_{\rm FB}}
\newcommand{\GeV}{{\,\mbox{GeV}}}
\newcommand{\MeV}{{\,\mbox{MeV}}}
\newcommand{\degree}{^\circ}
\newcommand{\mB}{m_B}
\newcommand{\SM}{{\rm SM}}
\newcommand{\NP}{{\rm NP}}
\newcommand{\barc}{\bar{c}}
\newcommand{\xipara}{\xi_\parallel^{\kone}}
\newcommand{\xiperp}{\xi_\perp^{\kone}}
\newcommand{\xiparal}{\xi_\parallel^{\konel}}
\newcommand{\xiperpl}{\xi_\perp^{\konel}}
\newcommand{\xiparah}{\xi_\parallel^{\koneh}}
\newcommand{\xiperph}{\xi_\perp^{\koneh}}
\newcommand{\para}{\parallel}
\newcommand{\alphas}{\alpha_s}
\newcommand{\pA}{p_{\kone}}
\newcommand{\lcaption}[2]{\caption{(label:{#2}) #1}\label{#2}}
\newcommand{\Rmunr}{R_{\mu,\rm nr}}
\newcommand{\RdGamma}{R_{d\Gamma/ds,\mu}}
\providecommand{\dfrac}[2]{\frac{\displaystyle
{#1}}{\displaystyle{#2}}}
\def\baeq{\begin{appeq}}     \def\eaeq{\end{appeq}}
\def\baeeq{\begin{appeeq}}   \def\eaeeq{\end{appeeq}}
\newenvironment{appeq}{\beq}{\eeq}
\newenvironment{appeeq}{\beeq}{\eeeq}
%
%
%
%
\title{\boldmath\bf
Form factors and decay rate of  $B_c^\ast \rightarrow D_{s}~l^+ l^-$ decays in the QCD sum rules}
\author{ K. Zeynali\\  Department of Sciences, Faculty of Medicine, Ardabil University of Medical Sciences, Ardabil, Iran\\( e-mail: k.zeinali@arums.ac.ir)\\
V. Bashiry\\
 Cyprus International University, Faculty of Engineering, Nicosia, Northern Cyprus, Mersin 10, Turkey\\
(e-mail: bahiry@ciu.edu.tr)\\F. Zolfagharpour\\
Department of Physics, University of Mohaghegh Ardabili, PO Box 179, Ardabil, Iran\\ (e-mail: zolfagharpour@uma.ac.ir)
 }

\setlength{\baselineskip}{24pt} \maketitle
\setlength{\baselineskip}{7mm}
\date{}
\maketitle \thispagestyle{empty}
Rare exclusive $B_c^\ast \rightarrow D_{s}~l^+ l^-$ decays are analyzed  in the framework of the
three--point QCD sum rules approach. The two  gluon condensate
corrections to the correlation function are included and the form factors of this transition are evaluated. Using the form factors, the  decay width and integrated decay rate for these decays are also calculated.
\newpage
\section{Introduction}
The rare flavor-changing neutral-current (FCNC) processes \{$b \to s(d)$\} are widely  studied  to test the predictions of Standard Model (SM) at loop level and to search for new-physics (NP). The  recent theoretical studies can be found in the Refs.\cite{Cogollo:2013mga}-\cite{Branco}.

Various physical observables of  leptonic, semileptonic and radiative $B$ decays have been measured by LHCb. For instance, the form factor, independent observables  in the decay $B^{0} \to K^{*0} \mu^+ \mu^-$\cite{Aaij:2013qta} and the  the $CP$ asymmetry in $B^+ \rightarrow K^+ \mu^+ \mu^-$ decays \cite{Aaij:2013dgw} have been measured. More recent measurements in the LHCb for FCNC transitions can be seen in Refs.\cite{Aaij:2013aka}-\cite{Aaij:2013iag}.  Measurements of various observables at LHCb indicate that SM predictions are in good agreement with the experimental results. Therefore, most of the new physics scenarios
 are excluded.

Rare \bcds proceeds FCNC transitions. This decay has not yet been measured by LHCb. There is not theoretical studies relevant to the form factors and decay rate of this decay. We try to calculate the form factors and the decay rate of \bcds  decay as well. We use the three--point QCD sum rules approach in the calculation of these form factors. The QCD sum rules have widely  been used to calculate form factors (some similar studies can be found in Refs.\cite{Bashiry:2013waa}-\cite{Marques de Carvalho:1999ia}).

The paper includes 3 sections: In section 2, we recall the effective Hamiltonian and use the
three--point QCD sum rules approach to calculate these form factors. In section 3, we will use the numerical values of form factors in order to determine the sensitivity of the decay rate to the invariant dileptonic mass and then present our conclusion.

\section{Sum rules for the  \bcds transition form factors}\label{sec:Hamiltonian}
The matrix element of the $b\rightarrow s\ell^{+}\ell^{-}$ transition can be written as:

\begin{eqnarray}
M(b \rightarrow s\ell ^{+}\ell ^{-})=\frac{G_{F}\alpha }{\sqrt{2}
\pi }V_{tb}V_{ts}^{\ast }\{ c_{9}^{eff}{\cal O}_9 +c_{10}{\cal O}_{10} -2\frac{{m}_{b}}{q^2}c_{7}^{eff}{\cal O}_7\}
 \label{e1}
\end{eqnarray}
where
 \begin{eqnarray}
 {\cal O}_7&=&\frac{1}{2}\left[ \bar{s}i\sigma _{\mu \nu }q^{\nu }(1+\gamma_5) b\right] \left[ \bar{\ell}\gamma^{\mu }\ell \right],\nnb\\
{\cal O}_9&=&\frac{1}{2}\left[ \bar{s}\gamma _{\mu }(1-\gamma_5)b\right] \left[\bar{\ell}\gamma^{\mu }\ell \right],\nnb\\
{\cal O}_{10}&=&\frac{1}{2}\left[ \bar{s}\gamma _{\mu }(1-\gamma_5)b\right] \left[ \bar{\ell}\gamma^{\mu}\gamma ^{5}\ell \right]\nnb,
 \end{eqnarray}
and $c_{7},~c_9$ and $c_{10}$ are Wilson coefficients evaluated  in the  naive dimensional regularization~(NDR) scheme at the leading order (LO),
next to leading order (NLO) and next-to-next leading order (NNLO) in
the SM\cite{Buras:1994dj}--\cite{NNLL}. $c_9^\eff(\hats) = c_9 +
Y(\hats)$, where $Y(\hats) = Y_{\rm pert}(\hats) + Y_{\rm LD}$
contains both the perturbative part $Y_{\rm pert}(\hats)$ and
long-distance part $Y_{\rm LD}(\hats)$.
$Y(\hats)_{\rm pert}$ in \cite{Buras:1994dj} is as follows:
\begin{eqnarray}
Y_{\rm pert} (\hats) &=& g(\hatm_c,\hats) c_0 \nonumber\\&&
-\frac{1}{2} g(1,\hats) (4 \barc_3 + 4 \barc_4 + 3 \barc_5 +
\barc_6) -\frac{1}{2} g(0,\hats) (\barc_3 + 3 \barc_4) \nonumber\\&&
+\frac{2}{9} (3 \barc_3 + \barc_4 + 3 \barc_5 + \barc_6),
\\
\mbox{where}\quad c_0 &\equiv& \barc_1 + 3\barc_2 + 3 \barc_3 +
\barc_4 + 3 \barc_5 + \barc_6,
\end{eqnarray}
and the function $g(x,y)$ is defined in \cite{Buras:1994dj}.  Here,
$\barc_1$ -- $\barc_6$ are the Wilson Coefficients in the leading
logarithmic approximation. The relevant Wilson Coefficients are
given  in \cite{Ali:1999mm}. $Y(\hats)_{\rm LD}$ involves $B_c^\ast
\to D_s V(\cbar c)$ resonances \cite{Lim:1988yu,
Ali:1991is,Kruger:1996cv}, where $V(\cbar c)$ are the vector
charmonium states. Following refs.~\cite{Lim:1988yu,Ali:1991is}, $Y_{\rm LD}(\hats)$ is as follows:

\begin{eqnarray}
Y_{\rm LD}(\hats) &=&
 - \frac{3\pi}{\alphaem^2} c_0
\sum_{V = \psi(1s),\cdots} \kappa_V \frac{\hatm_V \Br(V\to
\l^+\l^-)\hat{\Gamma}_{\rm tot}^V}{\hats - \hatm_V^2 + i \hatm_V
\hat{\Gamma}_{\rm tot}^V},
\end{eqnarray}
where $\hat{\Gamma}_{\rm tot}^V \equiv \Gamma_{\rm tot}^V/m_{B_c^\ast}$ and
$\kappa_V$ takes different value for different exclusive
semileptonic decays. This phenomenological parameters $\kappa_V$ can
be fixed for $B\rightarrow K^\ast \ell^+\ell^-$ decay by equating
the naive factorization estimate of the $B\rightarrow K^\ast V$ rate
and the results of the experimental measurements \cite{Ali:1999mm}. For the time being,  there is no experimental result on $B_c^\ast
\rightarrow D_s V (c\bar{c})$. Thus, we  use the results of
$B\rightarrow K^\ast V$ to estimate the values of $\kappa_V$ in our numerical calculations i.e, $\kappa_V=1.75$ for
$J/\Psi(1S)$ and $\kappa_V=2.43$ for $\Psi(2S)$, respectively.

Relevant properties of vector charmonium states are summarized in
Table~\ref{charmonium}.

\begin{table}[tbp]
\caption{Masses, total decay widths and branching fractions of
dilepton decays of vector charmonium states
\cite{pdg12}.}\label{charmonium}
\begin{center}
\begin{tabular}{cclll}
$V$ & Mass[\GeV] &  $\Gamma_{\rm tot}^V$[\MeV]
 &\multicolumn{2}{c}{$\Br(V\to\lpm)$}
\\
\hline $J/\Psi(1S)$ & $3.097$ & $0.093$ & $5.9\times10^{-2}$ & for
$\l=e,\mu$
\\
$\Psi(2S)$   & $3.686$ & $0.303$ & $7.82\times10^{-3}$ & for $\l=e,\mu$ \\
             &         &             & $3.0\times10^{-3}$ & for $\l=\tau$
\\
$\Psi(3770)$ & $3.773$ & $27.2$ & $9.6\times10^{-6}$ & for $\l=e$
\\
$\Psi(4040)$ & $4.040$ & $80$ & $1.1\times10^{-5}$ & for $\l=e$
\\
$\Psi(4160)$ & $4.153$ & $103$ & $8.1\times10^{-6}$ & for $\l=e$
\\
$\Psi(4415)$ & $4.421$ & $62$ & $9.4\times10^{-6}$ & for $\l=e$
\end{tabular}
\end{center}
\end{table}

 The transition amplitude  of the exclusive \bcds  decays is obtained by
sandwiching Eq.(\ref{e1}) between the initial meson state \bcs  and
the final meson state $D_s$ in terms of form factors as follows:

\begin{eqnarray}\label{KZVBFZ}
\nonumber M &=& \frac{G_{F}\alpha}{2\sqrt{2}\pi}
V_{tb}V_{ts}^{*}\Bigg[ c_9^{eff}
<D_{s}(p_D)\mid\overline{s} \gamma_\mu (1-\gamma_5)
b\mid B_{c}^\ast(p_B,\varepsilon)>  \overline{\ell}\gamma_\mu \ell \nnb\\&+& c_{10}
<D_{s}(p_D)\mid\overline{s} \gamma_\mu
(1-\gamma_5) b\mid B_{c}^\ast(p_B,\varepsilon)> \overline{\ell} \gamma_\mu \gamma_5 \ell \\ &-& 2 c_7^{eff}\frac{m_b}{q^2} <D_s(p_D)\mid
\overline{s}  ~i \sigma_{\mu\nu} q^\nu (1+\gamma_5) b\mid
B_{c}^\ast(p_B,\varepsilon)> \overline{l} \gamma_\mu l \Bigg]\nnb,
\end{eqnarray}
where $\varepsilon$ is the polarization vector of
\bcs  meson, $p_B$ and $p_D$ are  momentums  of the $B_c^\ast$ and $D_s$ mesons,
respectively. Taking the Lorentz invariance and
parity conservation into account, the  matrix elements of the Eq.(\ref{KZVBFZ}) are
parameterized in terms of the form factors as follows:
\begin{eqnarray}\label{3au}
\nonumber <D_{s}(p_D)\mid\overline{s}\gamma_{\mu}(1-\gamma_{5}) b\mid
B_c^\ast(p_B,\varepsilon)>&=&A_{V}(q^2)\varepsilon_{\mu\nu\alpha\beta}
\varepsilon^{\ast\nu}p_B^\alpha p_D^\beta-iA_{0}(q^2)\varepsilon_{\mu}^{\ast} \\
-iA_{+}(q^2)(\varepsilon^{*}p_D)P_{\mu}
&-&i A_{-}(q^2)(\varepsilon^{*}p_D)q_{\mu},
\end{eqnarray}
\begin{eqnarray}\label{4au}
\nonumber &<D_{s}(p_D)\mid\overline
{s}\sigma_{\mu\nu} q^\nu (1+\gamma_5) b\mid
B_{c}^\ast(p_B,\varepsilon)>=-T_{V}(q^2)~i\varepsilon_{\mu\nu\alpha\beta}
\varepsilon^{\ast\nu}p_B^\alpha p_D^\beta&\\
&-T_{0}(q^2)\Bigg\{\varepsilon_{\mu}^{\ast}+\frac{(\varepsilon^{*}p_D)}
{(m_{B_{c}^\ast}^2-m_{D_{s}}^2)}P_{\mu}\Bigg\}
-T_{+}(q^2)(\varepsilon^{*}p_D)\Bigg\{q_{\mu}
-\frac{q^2}{m_{B_{c}^\ast}^2-m_{D_{s}}^2}P_{\mu}\Bigg\}&,
\end{eqnarray}
where $A_{V}(q^2)$, $A_{0}(q^2)$, $A_{+}(q^2)$ , $A_{-}(q^2)$,
$T_{V}(q^2)$, $T_{0}(q^2)$ and $T_{+}(q^2)$ are the transition form
factors. $P_{\mu}=(p_B+p_D)_{\mu}$ and $q_{\mu}=(p_B-p_D)_{\mu}$, here,
$q$ is transfer momentum or the momentum of the $Z$ boson (photon).

The transition amplitude in terms of the form factors is as follows:

\begin{eqnarray}\label{ampilitude}
\nonumber M &=& \frac{G_{F}\alpha}{2\sqrt{2}\pi}
V_{tb}V_{ts}^{*}\Bigg\{ \overline{\ell}\gamma_\mu \ell \bigg[  A_1 \varepsilon_{\mu\nu\alpha\beta}p_B^\alpha p_D^\beta \varepsilon^\nu-A_2(\varepsilon.p_D)P_\mu-A_3(\varepsilon.p_D)q_\mu-A_4\varepsilon_\mu\bigg]\nnb\\&+&  \overline{\ell}\gamma_\mu \gamma_5 \ell \bigg[  B_1 \varepsilon_{\mu\nu\alpha\beta}p_B^\alpha p_D^\beta \varepsilon^\nu-B_2(\varepsilon.p_D)P_\mu-B_3(\varepsilon.p_D)q_\mu-B_4\varepsilon_\mu\bigg]\bigg\},
\end{eqnarray}

where
\begin{eqnarray}
  A_1 &=& -I c_9 A_V-2 C_7 T_V m_b/q^2 \nnb\\
  A_2 &=& I c_9 A_{+}+\frac{2 c_7 m_b}{q^2(m_B^{\ast2}-m_D^2)}(T_0+q^2 T_-)\nnb \\
  A_3 &=&-I c_9 A_{-}-2 C_7 T_- m_b/q^2 \nnb  \\
 A_4 &=&I c_9 A_{0}-2 C_7 T_0 m_b/q^2  \nnb \\
  B_1&=&-IA_V c_{10},~~B_2=A_+c_{10},~~  B_3=A_-c_{10},~~ B_4=A_0 c_{10}
\end{eqnarray}
The decay rate for the \bcds  decay is obtained as follows:

\bea\frac{d\Gamma}{dq^2}&=&\frac { {G_f}^2 \alpha ^2{|V_ {tb} V {ts}|}^2} {6144 \pi ^5 {m_ {B_c^*}}^3} {\lambda^{1/2} (m_ {B_c^*}^2, m_ {D_s}^2,
       q^2)} {v}\Delta\eea
       where
       \bea v=\sqrt{1-\frac{4m_\ell^2}{q^2}},
       \eea
\bea
\Delta&=&\lambda (m_{B_c^*}^2, m_ {D_s}^2, q^2)\bigg\{-2 |
{A_1}|^2 (2{m_\ell}^2 +
       q^2) +
    2 |{B_1}|^2   (4
           {m_\ell}^2 -
       q^2)- \frac {6 |{B_3}|^2  {m_\ell}^2 q^2} {{m_{B_c^*}}^2}+\frac {12
Re[{B_3B_4^*}]  {m_\ell}^2} {{m_ {B_c^*}}^2} \nnb\\&-&\frac {12 Re[{B_2B_3^*}]   {m_\ell}^2
          ({m_ {B_c^*}}^2 - {m_ {D_s}}^2)} {{m_ {B_c^*}}^2}\bigg\}- \frac{1}{{m_ {B_c^*}}^2 q^2}\bigg\{ {|{A2}|^2 \lambda (m_ {B_c^*}^2, m_ {D_s}^2,
         q^2) ^2 (2 {m_\ell}^2 +
         q^2)} {{m_ {B_c^*}}^2 q^2}  \nnb\\&+&  {2Re[{A_2A_4^*}] (2{m_\ell}^2 +q^2) \bigg({m_ {B_c^*}}^6 - {m_ {B_c^*}}^4 (3 {m_ {D_s}}^2 +
            q^2) + {m_ {B_c^*}}^2 (3 {m_ {D_s}}^4 - 2 {m_ {D_s}}^2
                q^2 - q^4) - ({m_ {D_s}}^2 -
            q^2)^3\bigg)} \nnb\\& -& {|{A_4}|^2 (2 \
{m_\ell}^2 + q^2) \bigg({m_ {B_c^*}}^4 - 2
             {m_ {B_c^*}}^2 ({m_ {D_s}}^2 - 5 q^2) + ({m_ {D_s}}^2 -
            q^2)^2\bigg)}  -|{B_2}|^2 \lambda (m_ {B_c^*}^2, m_ {D_s}^2,q^2) \bigg({m_ {B_c^*}}^4 (2 {m_\ell}^2 + q^2)\nnb\\&-&
         2 {m_ {B_c^*}}^2 ({m_ {D_s}}^2
                (2 {m_\ell}^2 + q^2) - 4 {m_\ell}^2 q^2 +
            q^4) + {m_ {D_s}}^4 (2 {m_\ell}^2 +
            q^2) +{m_ {D_s}}^2 (8 {m_\ell}^2 q^2 - 2
                q^4) - 4 {m_\ell}^2 q^4 +
         q^6\bigg) \nnb\\& +& {2 Re[{B_2B_4^*}] \lambda (m_{B_c^*}^2, m_ {D_s}^2, q^2)  \bigg({m_{B_c^*}}^2 (2
                {m_\ell}^2 + q^2) - {m_ {D_s}}^2 (2 {m_\ell}^2 +
            q^2) - 4 {m_\ell}^2 q^2 +
         q^4\bigg)} \nnb\\&-&  {{|B_4|}^2 \bigg({m_{B_c^*}}^4
             (2 {m_\ell}^2 + q^2) -
         2 {m_ {B_c^*}}^2 ({m_ {D_s}}^2 (2 {m_\ell}^2 + q^2) +
            26 {m_\ell}^2 q^2 - 5
                q^4) + ({m_ {D_s}}^2 - q^2)^2 (2 {m_\ell}^2 +
            q^2)\bigg)}\bigg\}\eea
 We follow the QCD sum rules approach to calculate the aforementioned form factors.  The QCD sum
rules start with the following correlation functions:
\begin{eqnarray}\label{6au}
{\cal{T}} _{\mu\nu}^{V-AV}(p_B^2,p_D^2,q^2)&=&i^2\int
d^{4}xd^4ye^{-ip_Bx}e^{ip_Dy}<0\mid T[J_{D_{s}}(y)
J_{\mu}^{V-AV}(0) J_{\nu B_{c}^\ast}(x)]\mid  0>,\nonumber\\
{\cal{T}}_{\mu\nu}^{T-PT}(p_B^2,p_D^2,q^2)&=&i^2\int
d^{4}xd^4ye^{-ip_Bx}e^{ip_Dy}<0\mid T[J_{D_{s}}(y)J_{\mu}^{T-PT}(0) J_{\nu B_{c}^\ast}(x)]\mid 0>,
\end{eqnarray}
where $J _{ D_{s}}(y)=\overline{c}\gamma_{5} s$ and $J_{\nu B_{c}^\ast}(x)=\overline{b}\gamma_{\nu}c$   are interpolating
currents of the  $D_s$ and \bcs ~ meson states, respectively.
 $J_{\mu}^{V-AV}=~\overline {s}\gamma_{\mu}(1-\gamma_{5})b $ ~and~
 $J_{\mu}^{T-PT}=~\overline {s}{\sigma_{\mu\nu} q^\nu (1+\gamma_5)}b$  consist of the  vector ($V$), axial vector ($AV$), tensor ($T$) and pseudo tensor ($PT$) transition currents.
The above correlation functions  can be re-written by inserting the two complete sets of the \bcs ~and $D_s$ meson currents with the same quantum numbers into the Eq.~(\ref{6au}) as follows:

\begin{eqnarray} \label{7au}
&&{\cal{T}} _{\mu\nu}^{V-AV}(p_B^2,p_D^2,q^2)=-
\nonumber \\
&&\frac{<0\mid J_{D_{s}} \mid
D_{s}(p_D)><D_{s}(p_D)\mid
J_{\mu}^{V-AV}\mid B_{c}^\ast(p_B,{\varepsilon})><B_{c}^\ast(p_B,{\varepsilon})\mid J_{\nu B_c^\ast}\mid
0>}{(p_D^2-m_{D_{s}}^2)(p_B^2-m_{B_c^\ast}^2)}+\cdots,
\nonumber \\
&&{\cal{T}} _{\mu\nu}^{T-PT}(p_B^2,p_D^2,q^2)=-
\nonumber \\
&&\frac{<0\mid J_{D_{s}} \mid
D_{s}(p_D)><D_{s}(p_D)\mid
J_{\mu}^{T-PT}\mid B_{c}^\ast(p_B,{\varepsilon})><B_{c}^\ast(p_B,{\varepsilon})\mid J_{\nu B_c^\ast}\mid
0>}{(p_D^2-m_{D_{s}}^2)(p_B^2-m_{B_c^\ast}^2)}+\cdots,
\end{eqnarray}
where "$\cdots$" indicates higher states and continuum contributions.
The $<0\mid J_{D_{s}} \mid
D_{s}(p_D)>$ and $<B_{c}^\ast(p_B,{\varepsilon})\mid J_{\nu B_c^\ast}\mid
0>$ matrix  elements are as follows:
\begin{equation}\label{8au}
 <0\mid J_{D_{s}} \mid
D_{s}(p_D)>=-i\frac{f_{D_{s}}m_{D_{s}}^2}{m_{s}+m_{c}} ~,~~<B_{c}^\ast(p_B,{\varepsilon})\mid J_{\nu B_c^\ast}\mid
0>=f_{B_{c}^\ast}m_{B_{c}^\ast}\varepsilon_\nu,
\end{equation}
where $f_{D_s}$ and $f_{B_{c}}$  are the leptonic decay
constants of the $D_{s} $ and $B_{c}^\ast$ mesons, respectively. Using
Eq.(\ref{3au}), Eq.(\ref{4au}) and Eq.(\ref{8au}) together with the
summation over the polarization of the $B_c^\ast$ meson, the
Eq.(\ref{7au}) can be re-written as follows:
\begin{eqnarray}\label{9amplitude}
{\cal{T}}_{\mu\nu}^{V-AV}(p_B^2,p_D^2,q^2)&=&-\frac{f_{D_{s}}m_{D_{s}}^2}
{(m_{c}+m_{s})}\frac{f_{B_{c}^{\ast}}m_{B_{c}^\ast}}
{(p_D^2-m_{D_{s}}^2)(p_B^2-m_{B_c^\ast}^2)} \times
\left[\vphantom{\int_0^{x_2}}A_{0}(q^2)g_{\mu\nu}+A_{+}(q^2)P_{\mu}p_{B\nu}\right.
\nonumber
\\ &+&\left. A_{-}(q^2)q_{\mu}p_{B\nu}+i\varepsilon_{\mu\nu\alpha\beta}
p_B^{\alpha}p_D^{\beta}A_{V}(q^2)\vphantom{\int_0^{x_2}}\right] + \mbox{excited states,}\nonumber\\
{\cal{T}}_{\mu\nu}^{T-PT}(p_B^2,p_D^2,q^2)&=&-\frac{f_{D_{s}}m_{D_{s}}^2}
{(m_{c}+m_{s})}\frac{f_{B_{c}^{\ast}}m_{B_{c}^\ast}}
{(p_D^2-m_{D_{s}}^2)(p_B^2-m_{B_c^\ast}^2)} \times
\left[\vphantom{\int_0^{x_2}}-i~T_{0}(q^2)g_{\mu\nu}\right.
\nonumber\\
&-&
\left.i~T_{+}(q^2)q_{\mu}p_{B\nu}+\varepsilon_{\mu\nu\alpha\beta}p_B^{\alpha}p_D^{\beta}T_{V}(q^2)\vphantom{\int_0^{x_2}}\right]
+ \mbox{excited states.}
\end{eqnarray}
Then, on  the QCD or theoretical side, the correlation functions are evaluated in terms of the  quarks and gluons parameters by means of the  the operator product expansion (OPE). The
correlation functions are written as:
\begin{eqnarray}\label{QCD side}
{\cal{T}}_{\mu\nu}^{V-AV}(p_B^2,p_D^2,q^2)&=&{\cal{T}}^{V-AV}_{0}g_{\mu\nu}
+{\cal{T}}^{V-AV}_{+}P_{\mu}p_{B\nu}+{\cal{T}}^{V-AV}_{-}q_{\mu}p_{B\nu}+
i{\cal{T}}^{V-AV}_{V}\varepsilon_{\mu\nu\alpha\beta}p_B^{\alpha}p_D^{\beta},
\nonumber\\
{\cal{T}}_{\mu\nu}^{T-PT}(p_B^2,p_D^2,q^2)&=&-i~{\cal{T}}^{T-PT}_{0}g_{\mu\nu}
-i~{\cal{T}}^{T-PT}_{+}q_{\mu}p_{B\nu}+{\cal{T}}^{T-PT}_{V}
\varepsilon_{\mu\nu\alpha\beta}p_B^{\alpha}p_D^{\beta},
\end{eqnarray}
where each ${\cal{T}}_{i}$ with $i=0, +, -$ and $V$ contains the perturbative and nonpertubative parts as:
\begin{eqnarray}\label{QCD side1}
{\cal{T}}_{i}&=&{\cal{T}}_{i}^{pert}+ {\cal{T}}_{i}^{nonpert}.
\end{eqnarray}
The perturbative part of the Eq.(\ref{QCD side1}) is  the bare-loop diagram given in Fig.1(a). The nonperturbative part consists of the light quark condensates and the two gluon condensates diagrams \{see Fig.2(a-f)\}. Contributions of
the light quark condensates \{diagrams shown in Fig.1(b, c, d)\} are removed  by the application of the
double Borel transformations \cite{Azizi:2008vv}. Therefore, the two gluon condensates diagrams shown in Fig.2(a-f) are described as the first correction.
\begin{figure}
  \vspace*{1cm}
  \centering
  \includegraphics[width=11cm]{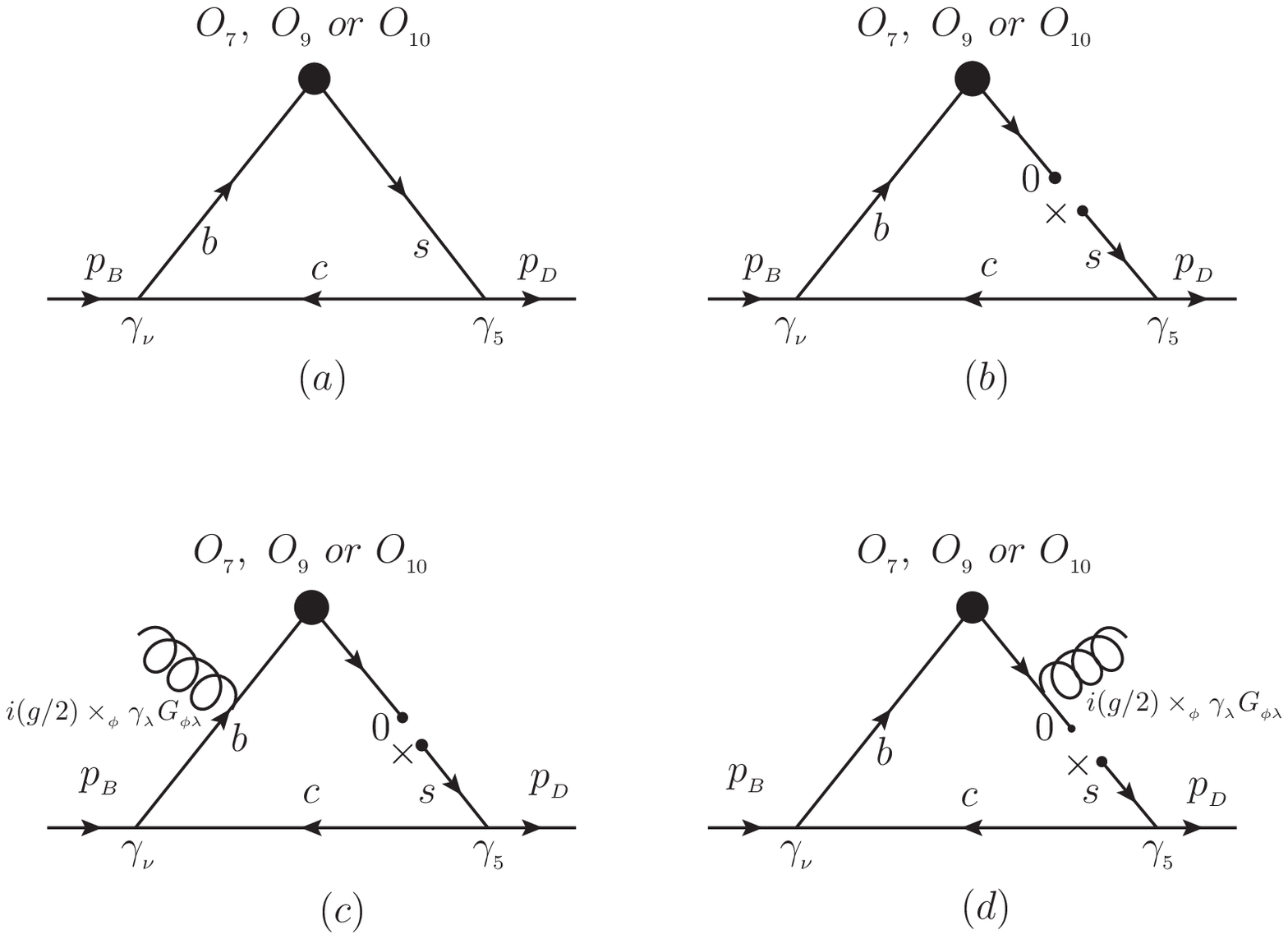}\\
  \caption{The bare-loop and light quarks condensates contributions to $B_c^\ast \rightarrow
D_{s}~l^+ l^-$ transitions}\label{fig1}
\end{figure}

 The bare-loop contributions for each structure in the correlation function are the double dispersion integrals given in the following formula:

\begin{equation}\label{10au}
{\cal{T}}_i^{per}=-\frac{1}{(2\pi)^2}\int du\int
ds\frac{\rho_{i}(s,u,q^2)}{(s-p_B^2)(u-p_D^2)}+\textrm{ subtraction
terms}.
\end{equation}
In the foregoing calculations of the spectral density $\rho_{i}(s,u,q^2)$,  Cutkosky Rules are applied to the Feynman Integrals. According to these rules   the quark propagators are replaced by Dirac Delta Functions:
$\frac{1}{p^2-m^2}\rightarrow-2\pi i\delta(p^2-m^2),$ which means that all quarks are real.

The integration region in Eq.(\ref{10au}) is restricted by the arguments of the three $\delta$ functions in $s, u$ and $q^2$ space coordinates, where these $\delta$ functions must be zero at the same time. As a result, the following inequality in $s,u$ and $q^2$ space coordinates is obtained:
 \begin{equation}\label{13au}
 -1\leq\frac{2su+(s+u-q^2)(m_{b}^2-s-m_{c}^2)+(m_{c}^2-m_{s}^2)2s}
 {\lambda^{1/2}(m_{b}^2,s,m_{c}^2)\lambda^{1/2}(s,u,q^2)}\leq+1
\end{equation}
where $ \lambda(a,b,c)=a^2+b^2+c^2-2ac-2bc-2ab$.

Following the required calculations, the spectral densities are obtained as:

\bea \rho^{V-AV}_V&=&N_c I_0(s,u,q^2)\Bigg\{C_1
(m_b-m_c)-(C_2+1) m_c+C_2 m_s\Bigg\}\nnb\\
 \rho^{V-AV}_0 &=&\frac{N_c}{2} I_0(s,u,q^2)\Bigg\{-2
m_c^3+2 m_s m_c^2-[(C_1+C_2+1) (-q^2+s+u)+2 C_1 s
 \nnb \\ &+&
2 C_2 u] m_c+ m_b[2 m_c^2-2 m_s m_c+2 C_2u+C_1 (-q^2+s+u)]+m_s
[2 C_1 s
\nnb\\ &+& C_2 (-q^2+s+u)] \Bigg\}\nnb\\
 \rho^{V-AV}_{+}&=&\frac{N_c}{2} I_0(s,u,q^2)\Bigg\{
C_1(m_b-2C_2m_c-m_c+2C_2 m_s)\nnb\\&-&
(2 C_2+1) (C_2m_c+m_c-C_2 m_s) \Bigg\}\nnb\\
 \rho^{V-AV}_{-}&=&\frac{N_c}{2} I_0(s,u,q^2)\Bigg\{(2C_2-1) (C_2 m_c+m_c-C_2 m_s)
\nnb\\&+&C_1 (m_b-2 C_2m_c-m_c+2 C_2 m_s) \Bigg\}\nnb\\
\rho^{T-PT}_V&=&4 N_c I_0(s,u,q^2)\Bigg\{
-2 s C_1^2-(m_c^2-m_s m_c+s+2 C_2 (-q^2+s+u)) C_1 \nnb \\
&-& C_2 m_c^2+C_2 m_c m_s+m_c m_s+m_b ((C_1+C_2+1) m_c-(C_1+C_2)
m_s)-2 C_2^2 u \nnb \\ &-& C_2 u \Bigg\}\nnb\\
\rho^{T-PT}_0&=&2 N_c I_0(s,u,q^2)\Bigg\{ C_2
q^2 m_c^2-C_2s m_c^2-2 s m_c^2+C_2 u m_c^2+2 u m_c^2
   \nnb \\ &-&
   C_2 m_s q^2 m_c+m_s q^2m_c+C_2 m_s s m_c+m_s s m_c-C_2 m_s u m_c-m_s um_c+C_2 u^2
   \nnb \\ &-&
   C_2 q^2 u-C_2 s u+C_1 [-(q^2+s-u)m_c^2+m_s (q^2+s-u) m_c+s (q^2-s+u)]
   \nnb \\ &+&
   m_b [m_c((C_1-C_2-1) q^2+(C_1+C_2+1) (s-u))+m_s (C_2(q^2-s+u)
   \nnb \\ &-&
   C_1 (q^2+s-u))]\Bigg\}\nnb\\
 \rho^{T-PT}_{+} &=& 2 N_c I_0(s,u,q^2)\Bigg\{((C_1+C_2+2) m_c^2-(C_1+C_2+1) m_s m_c+2 C_1 C_2 q^2
\nnb\\ &-&  m_b ((C_1+C_2+1) m_c-(C_1+C_2) m_s)+C_1 s+C_2 u)
\Bigg\}\eea

where
\bea
I_{0}(s,u,q^2)&=&\frac{1}{4\lambda^{1/2}(s,u,q^2)},\nonumber\\
C_1&=&\frac{m_c^2 (s - u - q^2) + u (2 m_b^2 - s + u - q^2) - m_s^2 (s + u - q^2)}{\lambda(s,u,q^2)}\nnb\\
C_2&=&\frac{s (2 m_s^2 + s - u - q^2) - m_b^2 (s + u - q^2) - m_c^2 (s - u + q^2)}{\lambda(s,u,q^2)}\nnb\\
N_c&=&3.
\eea

Now, the aim here is to evaluate the nonperturbative part of the Eq.(\ref{QCD side1}). As has already been mentioned, the contributions of the  the light quark condensate diagrams (Fig. 1b,1c and 1d) to the nonperturbative part of the correlation function vanish\cite{Azizi:2008vv}.
Thus, the gluon condensates diagrams shown in Fig.2 are evaluated.
The Fock--Schwinger fixed--point gauge
\cite{R7320,R7321,R7322} is used, where $x^\mu A_\mu^a = 0~,$
$A_\mu^a$ is the gluon field.

The following  integrals must be solved while evaluating the gluon condensate diagrams:
\cite{Aliev:2006vs,R7323}: \bea \label{e7323} I_0[a,b,c] = \int
\frac{d^4k}{(2 \pi)^4} \frac{1}{\left[ k^2-m_b^2 \right]^a \left[
(p_B+k)^2-m_c^2 \right]^b \left[ (p_D+k)^2-m_s^2\right]^c}~,
\nnb \\ \nnb \\
I_\mu[a,b,c] = \int \frac{d^4k}{(2 \pi)^4} \frac{k_\mu}{\left[
k^2-m_b^2 \right]^a \left[ (p_B+k)^2-m_c^2 \right]^b \left[
(p_D+k)^2-m_s^2\right]^c}~,
\nnb \\ \nnb \\
I_{\mu\nu}[a,b,c] =\int \frac{d^4k}{(2 \pi)^4} \frac{k_\mu
k_\nu}{\left[ k^2-m_b^2 \right]^a \left[ (p_B+k)^2-m_c^2 \right]^b
\left[ (p_D+k)^2-m_s^2\right]^c}~, \eea where $k$ is the
momentum of the spectator quark $c$.

The integrals are transferred from Minkowski space--time to Euclidean space--time.
Then, the Schwinger representation for the Euclidean propagator is used as in the  following: \bea
\label{e7324} \frac{1}{k^2+m^2} = \frac{1}{\Gamma(\alpha)}
\int_0^\infty d\alpha \, \alpha^{n-1} e^{-\alpha(k^2+m^2)}~. \eea
The Eq.(\ref{e7324}) is convenient to perform the Borel transformation, that is: \bea
\label{e7325} {\cal B}_{\hat{p}^2} (M^2) e^{-\alpha p^2} = \delta
(1/M^2-\alpha)~. \eea
 Performing integration over loop momentum $k$ and auxiliary parameters used in the exponential
representation of propagators \cite{R7321}, and applying double
Borel transformations over $p_B^2$ and $p^{ 2}_D$, the
transformed form of the integrals (see
also \cite{R7321}) in Eq.(\ref{e7323}) can be written as:
\begin{figure}
\vspace*{1cm}
\begin{center}
\includegraphics[width=11cm]{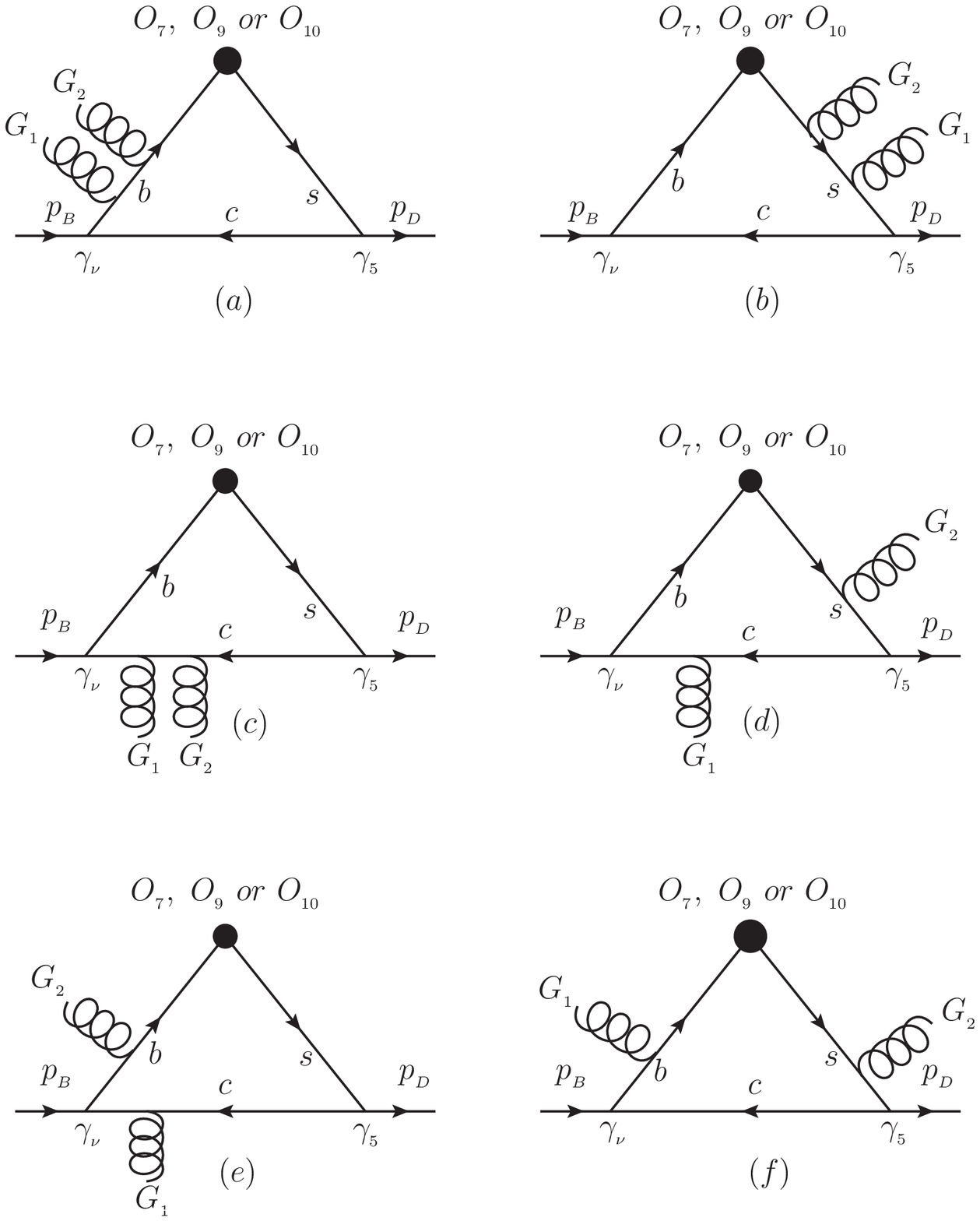}
\end{center}
\caption{Gluon condensate contributions to $B_c^* \rightarrow
D_{s}~l^+ l^-$ transitions  } \label{fig2}
\end{figure}
 \bea
\label{e7326} \hat{I}_0(a,b,c)& =&i\frac{(-1)^{a+b+c}}{16
\pi^2\,\Gamma(a) \Gamma(b) \Gamma(c)}
(M_1^2)^{2-a-b} (M_2^2)^{2-a-c} \, {\cal U}_0(a+b+c-4,1-c-b)~, \nnb \\ \nnb \\
\hat{I}_\mu(a,b,c) &=&\hat{I}_1(a,b,c) p_\mu +
\hat{I}_2(a,b,c) p'_\mu~, \nnb \\ \nnb \\
\hat{I}_{\mu\nu}(a,b,c)& =& \hat{I}_3(a,b,c) g_{\mu\nu} +
 \hat{I}_4(a,b,c)
p_\mu p_\nu
+ \hat{I}_5(a,b,c)  p'_\mu p'_\nu \nnb \\
&+&  \hat{I}_6 (a,b,c)  p_\mu p'_\nu + \hat{I}_7 (a,b,c)  p_\nu
p'_\mu, \eea where
\begin{eqnarray}
 \hat{I}_1(a,b,c) &=& i \frac{(-1)^{a+b+c+1}}{16
\pi^2\,\Gamma(a) \Gamma(b) \Gamma(c)}
(M_1^2)^{2-a-b} (M_2^2)^{3-a-c} \, {\cal U}_0(a+b+c-5,1-c-b)~, \nonumber \\ \nonumber \\
\hat{I}_2(a,b,c) &=& i \frac{(-1)^{a+b+c+1}}{16 \pi^2\,\Gamma(a)
\Gamma(b) \Gamma(c)}
(M_1^2)^{3-a-b} (M_2^2)^{2-a-c} \, {\cal U}_0(a+b+c-5,1-c-b)~, \nonumber \\ \nonumber \\
\hat{I}_3(a,b,c) &=& i \frac{(-1)^{a+b+c+1}}{32 \pi^2\,\Gamma(a)
\Gamma(b) \Gamma(c)}
(M_1^2)^{3-a-b} (M_2^2)^{3-a-c} \, {\cal U}_0(a+b+c-6,2-c-b)~,\nonumber \\ \nonumber \\
\hat{I}_4(a,b,c) &=& i \frac{(-1)^{a+b+c}}{16 \pi^2\,\Gamma(a)
\Gamma(b) \Gamma(c)}
(M_1^2)^{2-a-b} (M_2^2)^{4-a-c} \, {\cal U}_0(a+b+c-6,1-c-b)~,\nonumber \\ \nonumber \\
\hat{I}_5(a,b,c) &=& i \frac{(-1)^{a+b+c}}{16 \pi^2\,\Gamma(a)
\Gamma(b) \Gamma(c)}
(M_1^2)^{4-a-b} (M_2^2)^{2-a-c} \, {\cal U}_0(a+b+c-6,1-c-b)~,\nonumber \\ \nonumber \\
\hat{I}_6(a,b,c) &=& i \frac{(-1)^{a+b+c}}{16 \pi^2\,\Gamma(a)
\Gamma(b) \Gamma(c)} (M_1^2)^{3-a-b} (M_2^2)^{3-a-c} \, {\cal
U}_0(a+b+c-6,1-c-b)~,\nonumber \\ \nonumber \\
\hat{I}_7(a,b,c) &=&\hat{I}_6(a,b,c)~,
  \end{eqnarray}
 where $M_1^2$ and $M_2^2$ are the Borel
parameters.

The function ${\cal U}_0(\alpha,\beta)$ is:
\bea {\cal U}_0(a,b) = \int_0^\infty dy (y+M_1^2+M_2^2)^a y^b
\,exp\left[ -\frac{B_{-1}}{y} - B_0 - B_1 y \right]~, \nnb \eea
where \bea \label{e7328} B_{-1}& =& \frac{1}{M_1^2M_2^2}
\left[m_s^2M_1^4+m_b^2 M_2^4 + M_2^2M_1^2 (m_b^2+m_s^2
-q^2) \right] ~, \nnb \\
B_0 &=& \frac{1}{M_1^2 M_2^2} \left[ (m_s^2+m_c^2) M_1^2 + M_2^2
(m_b^2+m_c^2)
\right] ~, \nnb \\
B_{1} &=& \frac{m_c^2}{M_1^2 M_2^2}~. \eea

The Borel transformed form of the phenomenological side \{Eq.~(\ref{9amplitude})\}  and QCD side \{Eq.~(\ref{QCD side})\} is calculated.
The QCD sum rules for the form factors ( $A_{V}$, $A_{0}$,
$A_{+}$ , $A_{-}$, $T_{V}$, $T_{0}$ and $T_{-}$) can be achieved
by equating the  expressions of the phenomenological side \{Eq.~(\ref{9amplitude})\}  and QCD side \{Eq.~(\ref{QCD side})\} just after the following  Borel transformation:

\begin{eqnarray}\label{15au}
A_{i}(q^2)&=&\frac{(m_{s}+m_{c})e^{m_{B_{c}^*}^2/M_{1}^2}e^{m_{D_s}^2/M_{2}^2}
}{f_{B_{c}^*}m_{B_{c}^*}f_{D_s}m_{D_s}^2}\left[\vphantom{\int_0^{x_2}}\frac{1}{(2\pi)^2}\int_{u_{min}}^{u_0}
du
 \int_{s_{min}}^{s_0} ds\rho_{i}^{V-AV}(s,u,q^2)e^{-s/M_{1}^2-u/M_{2}^2}\right.\nonumber
\\&+&\left.i\frac{1}{24\pi^2}{C^{A_{i}}}<\frac{\alpha_{s}}{\pi}G^{2}>\vphantom{\int_0^{x_2}}\right],\\
T_{i}(q^2)&=&\frac{(m_{s}+m_{c})e^{m_{B_{c}^*}^2/M_{1}^2}e^{m_{D_s}^2/M_{2}^2}
}{f_{B_{c}^*}m_{B_{c}^*}^2f_{D_s}m_{D_s}^2}\left[\vphantom{\int_0^{x_2}}\frac{1}{(2\pi)^2}\int_{u_{min}}^{u_0}
du
 \int_{s_{min}}^{s_0} ds\rho_{i}^{T-PT}(s,u,q^2)e^{-s/M_{1}^2-u/M_{2}^2}\right.\nonumber
\\&+&\left.i\frac{1}{24\pi^2}{C^{T_{i}}}<\frac{\alpha_{s}}{\pi}G^{2}>\vphantom{\int_0^{x_2}}\right],
 \end{eqnarray}
Note that, the Borel transformation suppresses the contributions of higher states and continuum. In addition, the two gluon
condensates contributions are  $C^{A_{i}}$ and $C^{T_{i}}$. The contributions of the aforementioned expressions ($C^{A_{i}}$ and $C^{T_{i}}$) are considered in the numerical analysis. However, since each of these explicit expressions is extremely long, it is found unnecessary to show them all in this study. Therefore, one of these expressions ($C^{A_{V}}$) is shown as a sample in Appendix.
The $s_{0}$ and $u_{0}$ are the continuum
thresholds in $s$ and $u$ channels, respectively. Also $s_{min}=(m_b+m_c)^2$ and $u_{min}=(m_s+m_c)^2$.

\section{Numerical analysis}
From the explicit expressions for  the decay rate,
 it is clear that the main input parameters entering into
the expressions are  Wilson coefficients
$c_7^{eff}$, $c_9^{eff}$ and $c_{10}$ , the CKM matrix elements
 $\mid V_{tb}\mid=0.77^{+0.18}_{-0.24}$, $\mid
V_{ts}\mid=(40.6\pm2.7)\times10^{-3}$ \cite {pdg12}  and  the form factors  $A_{V}$,
$A_{0}$, $A_{+}$ , $A_{-}$, $T_{V}$, $T_{0}$, $T_{-}$. Moreover, the value of the gluon condensate $<\frac{\alpha_{s}}{\pi}G^{2}>=0.012~ GeV ^{4}$
\cite{Shifman1}, the masses and
leptonic decay constants, $m_{B_c^\ast}=6.2745\pm 0.0018$ GeV\cite{pdg12}, $m_{D_s}=1968.50\pm0.32 $ Mev\cite{pdg12},
$f_{B_{C}^{\ast}}$ and $f_{D_{s}}= (206.7 \pm 8.5 \pm2.5)$MeV\cite{pdg12}, the masses of the quarks
 $ m_{c}(\mu=m_{c})= 1.275\pm 0.015~ GeV$, $m_{s}(2~ GeV)\simeq 95 ~MeV$ \cite{pdg12} , and $m_{b} =(4.18\pm 0.03)~GeV$ \cite{pdg12}  are necessary for evaluation of the form factors.

In order to reduce the theoretical uncertainties as well as the dependence on the input parameters  used in the calculation of the form factors  the same set of the  input parameters is utilized for  calculation of  the form factors and the decay constant at the same time. We obtain $f_{B_{C}^{\ast}}=428\pm 35$ MeV by using generic formula given in Ref. \cite{Bashiry:2011pp}. This result is in good agreement with the results of the Ref.\cite{Wang:2012kw}, which is $f_{B_{C}^{\ast}}=415\pm 31$ MeV.
Furthermore, the form factors  contain four auxiliary
parameters: the Borel mass squares $M_{1}^2$ and $M_{2}^2$ and the continuum
threshold $s_{0}$  and $u_{0}$. The physical quantities such as  form factors are supposed to be independent or  weakly dependent  on
these auxiliary parameters in the so called "working regions".

The upper bound of the "working region"
of $M_{1}^2$ and $M_{2}^2$ is chosen  in a way that the contribution of continuum
is less than that of the first resonance. The lower
bound of $M_{1,2}^{2}$ is fixed  so that the contributions proportional to the highest
power of $1/M_{1,2}^{2}$ are less than about $30^{0}/_{0}$ of  the contributions proportional to the
highest power of $M_{1,2}^{2}$. With the aforementioned conditions, we find the stable region for the form factor in the following intervals; $10~GeV^2\leq M_{1}^{2}\leq25~GeV^2 $ and $4~GeV^2\leq M_{2}^{2}\leq10~GeV^2 $.
The continuum thresholds, $s_0$ and $u_0$   are determined by  the mass of the
corresponding ground-state hadron. The value of the $s_0$ and $u_0$  must  be less than the
energy of the first excited states with the same quantum numbers. Hence, the following regions for the  $s_0$ and $u_0$  are used:
$(m_{B_c^\ast}+0.3)^2\leq s_0\leq(m_{B_c^\ast}+0.7)^2$  and $(m_{D_s}+0.3)^2\leq u_0\leq (m_{D_s}+0.7)^2$ .

 In order to estimate the decay width of \bcds decays,  the $q^2$
dependency of the form factors  $A_{V}$, $A_{0}$, $A_{+}$ , $A_{-}$,
$T_{V}$, $T_{+}$ and  $T_{0}$ in the whole physical region, $
4m_{l}^2 \leq q^2 \leq (m_{B_{c}^\ast} - m_{D_{s}})^2$, is required.

The detailed numerical analysis of the form factor depicts that the dependence of the form factors   fits into the following function:

\begin{eqnarray}\label{fitfunction}
F(q^2)=\frac{a}{1- q^2/m_{fit}^2}+\frac{b}{(1-q^2/m_{fit}^2)^2}
\end{eqnarray}

The parameters of the fit function are given in Table \ref{tabpi}:
\begin{table}[h]
\center
\begin{tabular}{|c|c|c|c|}
\hline
& $m_{fit}$ & $a$& $b$\\
\hline
$A_V(q^2)$ & $5.3\pm 1.2 $ & $ -0.025\pm 0.005$ &$ 0.046\pm 0.007$ \\
\hline
$A_0(q^2)$ & $6.99\pm 1.5 $ & $ -0.83\pm0.13$ &$ 1.25\pm0.18$ \\
\hline
$A_+(q^2)$ & $5.28\pm 1.2 $ & $ -0.025\pm 0.005$ &$ 0.047\pm 0.007$ \\
\hline
$A_-(q^2)$ & $5.28\pm 1.2 $ & $ -0.025\pm 0.005$ &$ 0.047\pm 0.007$  \\
\hline
$T_V(q^2)$ & $5.26\pm 1.2 $ & $ -0.11\pm 0.017$ &$ 0.21\pm 0.033$  \\
\hline
$T_0(q^2)$ &$6.91\pm 1.14 $ & $ -3.71\pm 0.48$ &$ 5.62\pm 1.13$  \\
\hline
$T_+(q^2)$ & $6.58\pm 1.08 $ & $ -0.019\pm 0.003$ &$ 0.029\pm 0.004$    \\
 \hline
\end{tabular}
\caption{Parameters appearing in the form factors of the
\bcds decay in a four-parameter fit, for $M_{1}^2=17~GeV^2$, $M^2_2= 17$ GeV$^2$, $s_0=46$GeV$^2$ and $u_0=6$GeV$^2$}\label{tabpi}
\end{table}

The errors in the numerical calculation shown in Table II stem from the variation of the continuum thresholds,  the Borel mass parameter in the given
intervals and the uncertainties of the input parameters.

\begin{figure}
  \vspace*{1cm}
  \centering
   \includegraphics[width=11cm]{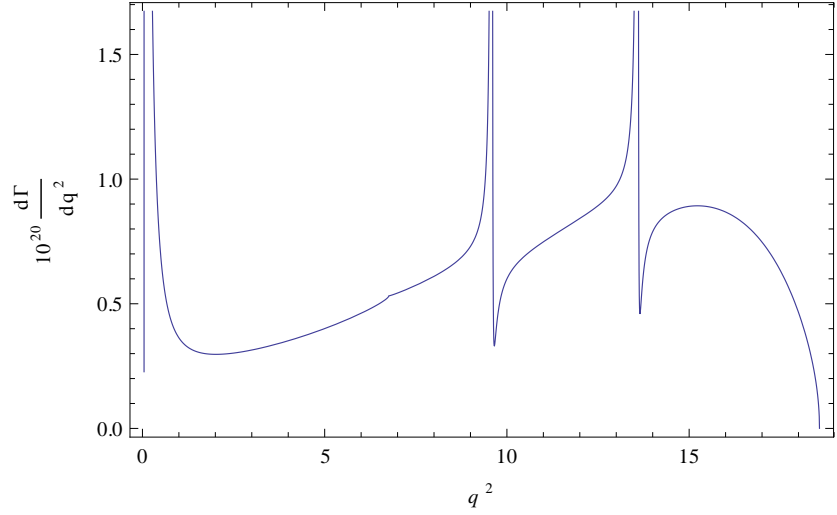}\\
  \caption{The dependence  of the differential decay rate on $q^2$  for $B_c^* \rightarrow
D_{s}~\mu^+ \mu^-$ transitions}\label{muplot}
\end{figure}

Taking  account of the dileptonic invariant mass( $q^{2}$) dependence of the form factors in the kinematical allowed region in the range of $4m_{l}^{2}\leq q^{2}\leq (m_{B_c^{\ast}}-m_{D_{s}} )^{2}$, we study the the differential decay rate for the \bcds ~decays. Our results for three different values of the $q^2$ are presented in Table \ref{branch}. In addition, Fig. (\ref{muplot}) depicts the dependence of the differential decay rate on $q^2$ for full kinematical allowed region.

The branching ratio can be calculated if we know the mean life time of the $B_c^\ast$. There is no experimental data on the mean life time of the $B_c^\ast$. Thus, we ignore about the calculation of the banting ratio. It is worth mentioning that using the theoretical methods like  Bethe-Salpeter model \cite{AbdElHady:1998kc} and potential model \cite{Kiselev:2000jc},  it is possible to calculate the mean life time of the $B_c^\ast$ meson and estimate the branching ratio as well.

\begin{table}[h]
\centering
\begin{tabular}{|c|c|}
 \hline
  $q^2($GeV$^2)$ & $(d\Gamma/dq^2)(B_{c}^\ast\rightarrow D_{s}\mu^{+}\mu^{-}$) \\
  \hline
$1$& $(3.63\pm 1.1)\times 10^{-21}$\\
\hline
$6$&$(4.61\pm 1.6)\times 10^{-21}$   \\
\hline
 $12$&$(8.42\pm 2.2)\times 10^{-21}$\\
 \hline
  \end{tabular}
\caption{Values for the decay rate of  the  $B_{c}^\ast\rightarrow
D_{s}\mu^+ \mu^-$ decay at three different values of the dileptonic invariant mass.} \label{branch}
\end{table}

Finally, we calculate   the integrated  decay rate
 for the $B_c^\ast \rightarrow D_{s}~\mu^+ \mu^-$ decays  as follows:
\begin{equation}\label{intbr}
  \Gamma=\int^{(m_{B_c^\ast}-m_{D_{s}} )^{2}}_{4m_\mu^2} \frac{d\Gamma}{dq^2}dq^2=(3.14\pm 0.82)\times10^{-19}
\end{equation}

To sum up, we investigated the $B_{c}^\ast\rightarrow D_{s}\ell^+\ell^-$
decays in the framework of the QCD sum rules approach. The form factors of these decays
were obtained in terms of the $q^2$. The  contributions of the quark
condensates in the correlations function found to be zero, so
 the contributions of the two gluon condensates to the correlations function were evaluated. Finally, we
calculated the differential decay width and the integrated decay rate of these decays for the muon channel.

\newpage


\newpage

\section*{Appendix}
In this section, we present the explicit expressions for the
coefficients $C^{A_{V}}$ corresponding to the
gluon condensates contributions of $g_{\mu\nu}$ structure entering to the expressions for the form factors
in Eq.(\ref{15au}).
\bea C^{A_{V}}&=&(8 m_b-16 m_s+16 m_c) I(1,1,2)+(24 m_s^3-40 m_b m_s^2+16 m_c m_s^2
  \nnb  \\ &+& 24 m_b^2 m_s+32 m_c^2 m_s-32 q^2 m_s-16 m_b m_c m_s-32 m_c^3-16 m_b m_c^2-8 m_c q^2) I(1,1,3)
   \\ \nnb &+&
   (-48 m_c^2 m_s^3+48 m_c^3 m_s^2+48 m_b m_c^2 m_s^2-48 m_b m_c^3 m_s+24 m_c^2 q^2 m_s) I(1,1,4)
    \\ \nnb &-&
   56 m_s I(1,2,1)+(8 m_s^3-8 m_b m_s^2+8 m_c^2 m_s-32 m_b m_c m_s-16 m_c^3-8 m_b m_c^2
    \\ \nnb &-&
   8 m_c q^2) I(1,2,2)+(24 m_s^5-24 m_b m_s^4-40 m_c m_s^4+8 m_c^2 m_s^3-16 q^2 m_s^3+32 m_b m_c m_s^3
    \\ \nnb &-&
   8 m_b m_c^2 m_s^2-8 m_c q^2 m_s^2+16 m_c^4 m_s+16 m_b m_c^3 m_s-8 m_c^2 q^2 m_s-8 m_c^5-16 m_b m_c^4
   \\ \nnb &-&
   8 m_c^3 q^2) I(1,2,3)+(-64 m_s^3+16 m_b m_s^2+16 m_c m_s^2+24 m_b^2m_s+24 m_c^2 m_s+8 q^2 m_s
   \\ \nnb &-&
   16 m_b m_c m_s-24 m_b m_c^2-24 m_b^2 m_c)I(1,3,1)+(16 m_s^5-16 m_b m_s^4-16 m_c m_s^4
   \\ \nnb &+&
   8 m_c^2 m_s^3-24 q^2 m_s^3+16 m_bm_c m_s^3-32 m_c^3 m_s^2-8 m_b m_c^2 m_s^2+24 m_c^4 m_s+32 m_b m_c^3 m_s
   \\ \nnb &-&
   16 m_c^2 q^2 m_s-24 m_b m_c^4) I(1,3,2)+(8 m_s^7-8 m_b m_s^6-16 m_c m_s^6-8m_c^2 m_s^5-8 q^2 m_s^5
   \\ \nnb &+&
   16 m_b m_c m_s^5+32 m_c^3 m_s^4+8 m_b m_c^2 m_s^4-8m_c^4 m_s^3-32 m_b m_c^3 m_s^3+16 m_c^2 q^2 m_s^3-16 m_c^5 m_s^2
   \\ \nnb &+&
   8 m_b m_c^4 m_s^2+8 m_c^6 m_s+16 m_b m_c^5 m_s-8 m_c^4 q^2 m_s-8 m_b m_c^6)I(1,3,3)+(-48 m_s^5
   \\ \nnb &+&
   48 m_b m_s^4+48 m_c m_s^4+24 q^2 m_s^3-48 m_b m_c m_s^3)I(1,4,1)+(8 m_c-8 m_s) I(2,1,1)
   \\ \nnb &+&
   (24 m_s m_b^2+16 m_s m_c m_b-8 m_s m_c^2-16m_s q^2) I(2,1,2)+(24 m_s m_b^4-24 m_s m_c^2 m_b^2
   \\ \nnb &-&
   16 m_s^2 m_c m_b^2+48m_s m_c^3 m_b-48 m_s^2 m_c^2 m_b+32 m_s^2 q^2 m_b-32 m_s m_c q^2 m_b+48 m_s m_c^4
   \\ \nnb &+&
   24 m_s q^4-32 m_s^2 m_c^3-88 m_s m_c^2 q^2+16 m_s^2 m_c q^2) I(2,1,3)+(-16m_b^3+16 m_s m_b^2
   \\ \nnb &-&
   8 m_c m_b^2-16 m_s^2 m_b-8 q^2 m_b+16 m_s m_c m_b+16 m_s^3-8m_s q^2-8 m_s^2 m_c) I(2,2,1)
   \\ \nnb &+&
   (16 m_s^5-16 m_b m_s^4-16 m_c m_s^4+8 m_b^2m_s^3-24 q^2 m_s^3+16 m_b m_c m_s^3-32 m_b^3 m_s^2
   \\ \nnb &-&
   8 m_b^2 m_c m_s^2+24 m_b^4m_s-16 m_b^2 q^2 m_s+32 m_b^3 m_c m_s-24 m_b^4 m_c) I(2,3,1)
   \\ \nnb &+&
   (-32 m_b^3+32m_s m_b^2-16 m_c m_b^2+16 m_s^2 m_b-8 q^2 m_b-16 m_s m_c m_b-72 m_s^3
   \\ \nnb &+&
   24 m_sm_c^2+16 m_s q^2+56 m_s^2 m_c) I(3,1,1)+(16 m_s m_b^4-32 m_s^2 m_b^3+16 m_s m_c m_b^3
    \\ \nnb &+&
   8 m_s m_c^2 m_b^2-40 m_s q^2 m_b^2-16 m_s^2 m_c m_b^2+32 m_s m_c^3 m_b-16 m_s^2 m_c^2 m_b+16 m_s^2 q^2 m_b
   \\ \nnb &-&
   32 m_s m_c q^2 m_b+24 m_s m_c^4+24 m_sq^4-32 m_s^2 m_c^3-48 m_s m_c^2 q^2+32 m_s^2 m_c q^2) I(3,1,2)
   \\ \nnb &+&
   (8 m_s m_b^6-16 m_s^2 m_b^5+16 m_s m_c m_b^5-8 m_s m_c^2 m_b^4-24 m_s q^2 m_b^4-16 m_s^2 m_c m_b^4
   \\ \nnb &-&
   32 m_s m_c^3 m_b^3+32 m_s^2 m_c^2 m_b^3+32 m_s^2 q^2 m_b^3-32 m_s m_c q^2m_b^3-8 m_s m_c^4 m_b^2+24 m_s q^4 m_b^2
   \\ \nnb &+&
   32 m_s^2 m_c^3 m_b^2-16 m_s m_c^2 q^2m_b^2+32 m_s^2 m_c q^2 m_b^2+16 m_s m_c^5 m_b-16 m_s^2 m_c^4 m_b
   \\ \nnb &-&
   16 m_s^2 q^4 m_b+16 m_s m_c q^4 m_b-32 m_s m_c^3 q^2 m_b+32 m_s^2 m_c^2 q^2 m_b+8 m_s m_c^6-8 m_s q^6
   \\ \nnb &-&
   16 m_s^2 m_c^5+24 m_s m_c^2 q^4-16 m_s^2 m_c q^4-24 m_s m_c^4 q^2+32 m_s^2 m_c^3 q^2) I(3,1,3)
   \\ \nnb &+&
   (-8 m_b^5+16 m_s m_b^4-16 m_c m_b^4-8 q^2 m_b^3+16 m_s m_cm_b^3+8 m_s^3 m_b^2-8 m_s q^2 m_b^2
   \\ \nnb &-&
   8 m_s^2 m_c m_b^2-40 m_s^4 m_b-8 m_s^2 q^2m_b+32 m_s^3 m_c m_b+24 m_s^5-16 m_s^3 q^2
   \\ \nnb &-&
   24 m_s^4 m_c) I(3,2,1)+(8 m_s^7-16m_b m_s^6-8 m_c m_s^6-8 m_b^2 m_s^5-8 q^2 m_s^5+16 m_b m_c m_s^5
   \\ \nnb &+&
   32 m_b^3m_s^4+8 m_b^2 m_c m_s^4-8 m_b^4 m_s^3+16 m_b^2 q^2 m_s^3-32 m_b^3 m_c m_s^3-16m_b^5 m_s^2
   \\ \nnb &+&
   8 m_b^4 m_c m_s^2+8 m_b^6 m_s-8 m_b^4 q^2 m_s+16 m_b^5 m_c m_s-8m_b^6 m_c) I(3,3,1)+(-144 m_s^2 m_b^3
   \\ \nnb &+&
   144 m_s m_c m_b^3+144 m_s^3 m_b^2-72m_s q^2 m_b^2-144 m_s^2 m_c m_b^2) I(4,1,1)
   \\ \nnb &+&
   (8 m_b q^2-8 m_c q^2)I_1(1,1,3)+(24 m_s m_c^2 q^2-24 m_b m_c^2 q^2) I_1(1,1,4)
   \\ \nnb &+&
   (8 m_b q^2+24 m_s q^2)I_1(1,3,1)+(24 m_s^3 q^2-24 m_b m_s^2 q^2) I_1(1,4,1)
   \\ \nnb &+&
   (8 m_b q^2-8 m_c q^2)I_1(2,1,2)+(-16 m_b q^4+8 m_c q^4-24 m_c^3 q^2+24 m_b m_c^2 q^2
   \\ \nnb &+&
   8 m_b^2 m_c q^2)I_1(2,1,3)+(16 q^2 m_b^3-8 m_s q^2 m_b^2+8 m_s^2 q^2 m_b-16 m_s^3 q^2) I_1(2,3,1)
   \\ \nnb &+&
   (40 m_s q^2-16 m_b q^2) I_1(3,1,1)+(-16 m_b q^4+24 m_b^3 q^2+16 m_b m_c^2 q^2)I_1(3,1,2)
   \\ \nnb &+&
   (8 m_b q^6-16 m_b^3 q^4-16 m_b m_c^2 q^4+8 m_b^5 q^2+8 m_b m_c^4 q^2-16 m_b^3m_c^2 q^2) I_1(3,1,3)
   \\ \nnb &+&
   (16 q^2 m_b^3-8 m_s q^2 m_b^2+8 m_s^2 q^2 m_b-16 m_s^3 q^2)I_1(3,2,1)+(8 q^2 m_b^5-8 m_s q^2 m_b^4
   \\ \nnb &-&
   16 m_s^2 q^2 m_b^3+16 m_s^3 q^2 m_b^2+8 m_s^4 q^2m_b-8 m_s^5 q^2) I_1(3,3,1)
   \\ \nnb &+&
   (72 m_b^3 q^2-72 m_b^2 m_s q^2) I_1(4,1,1)+(-8 m_sq^2-16 m_c q^2) I_2(1,1,3)
   \\ \nnb &+&
   (24 m_s m_c^2 q^2-24 m_c^3 q^2) I_2(1,1,4)+(8 m_s q^2-8m_c q^2) I_2(1,2,2)+(-16 q^2 m_s^3
   \\ \nnb &+&
   8 m_c q^2 m_s^2-8 m_c^2 q^2 m_s+16 m_c^3 q^2)I_2(1,2,3)+(24 m_s q^2+8 m_c q^2) I_2(1,3,1)
   \\ \nnb &+&
   (-16 q^2 m_s^3+8 m_c q^2 m_s^2-8 m_c^2q^2 m_s+16 m_c^3 q^2) I_2(1,3,2)+(-8 q^2 m_s^5+8 m_c q^2 m_s^4
   \\ \nnb &+&
   16 m_c^2 q^2 m_s^3-16m_c^3 q^2 m_s^2-8 m_c^4 q^2 m_s+8 m_c^5 q^2) I_2(1,3,3)
   \\ \nnb &+&
   (24 m_s^3 q^2-24 m_s^2 m_cq^2) I_2(1,4,1)+(8 m_c q^2-8 m_b q^2) I_2(2,1,2)
   \\ \nnb &+&
   (40 m_c^3 q^2-16 m_c q^4)I_2(2,1,3)+(-8 m_b q^2+48 m_s q^2-40 m_c q^2) I_2(3,1,1)
   \\ \nnb &+&
   (8 m_b q^4-16 m_c q^4-8 m_b^3q^2+16 m_c^3 q^2-8 m_b m_c^2 q^2+8 m_b^2 m_c q^2) I_2(3,1,2)
   \\ \nnb &+&
   (8 m_c q^6-16 m_c^3 q^4-16m_b^2 m_c q^4+8 m_c^5 q^2-16 m_b^2 m_c^3 q^2+8 m_b^4 m_c q^2) I_2(3,1,3)
   \\  &+&
   (72 m_b^2 m_c q^2-72 m_b^2 m_s q^2) I_2(4,1,1)
   \nnb\\&+&
   D_3^0 \Bigg\{8 m_s I(3,3,1)+(8 m_c-8 m_b)
I_1(3,3,1)\Bigg\} \nnb  \\ \nnb &+& D^3_0 \Bigg\{8 m_b I(1,3,3)+(8
m_c-8 m_s) I_2(1,3,3)\Bigg\}
\\ \nnb &+&
D_0^2 \Bigg\{(-24 m_b+8 m_c-16 m_s) I(1,2,3)+(8 m_c-24 m_b)I(1,3,2)
\\ \nnb &+&
(8 m_c^3-24 m_b m_c^2-16 m_s m_c^2+8 m_s^2 m_c-8 q^2 m_c+16 m_b m_s m_c-24 m_b m_s^2)I(1,3,3)
   \\ \nnb &+&
   (8 m_s-16 m_c) I_2(1,2,3)+(16 m_s-8 m_c) I_2(1,3,2)
   \\ \nnb &+&
   (-16 m_c^3+16 m_s m_c^2-16 m_s^2 m_c-8 q^2 m_c+16 m_s^3+8 m_s q^2) I_2(1,3,3)\Bigg\}
   \\ \nnb &+&
   D^0_2\Bigg\{ D^1_0\Bigg[8 m_c I(3,3,1)+(8 m_c-8 m_b) I_1(3,3,1)+(16 m_c-16 m_b) I_2(3,3,1)\Bigg]
   \\ \nnb &+&
   (8m_c-24 m_s) I(2,3,1)+(-16 m_b+8 m_c-24 m_s) I(3,2,1)+(8 m_c^3-16 m_b m_c^2
   \\ \nnb &-&
   24m_s m_c^2+8 m_b^2 m_c-8 q^2 m_c+16 m_b m_s m_c-24 m_b^2 m_s) I(3,3,1)
   \\ \nnb &+&
   (16m_b-8 m_c) I_1(2,3,1)+(8 m_b-16 m_c) I_1(3,2,1)
   \\ \nnb &+&
   (16 m_b^3-16 m_c m_b^2+16 m_c^2 m_b+8 q^2 m_b-16 m_c^3-8 m_c q^2) I_1(3,3,1)\Bigg\}
   \\ \nnb &+&
   D_0^1 \Bigg\{ D^2_0\Bigg[8 m_c I(1,3,3)+(16 m_c-16 m_s) I_1(1,3,3)+(8 m_c-8 m_s) I_2(1,3,3)\Bigg]
   \\ \nnb &+&
   D^1_0\Bigg[(-16 m_c-8 m_s) I(1,2,3)-16 m_c I(1,3,2)+(-16 m_c^3-16 m_s^2 m_c) I(1,3,3)
   \\ \nnb &-&
   16 m_c I(2,3,1)+(-8 m_b-16 m_c) I(3,2,1)+(-16 m_c^3-16 m_b^2 m_c) I(3,3,1)
   \\ \nnb &+&
   (16m_s-32 m_c) I_1(1,2,3)+(32 m_s-16 m_c) I_1(1,3,2)+(-32 m_c^3+32 m_s m_c^2
   \\ \nnb &-&
   32 m_s^2 m_c+32 m_s^3) I_1(1,3,3)+(16 m_b-8 m_c) I_1(2,3,1)+(8 m_b-16 m_c) I_1(3,2,1)
   \\ \nnb &+&
   (16 m_b^3-16 m_c m_b^2+16 m_c^2 m_b-16 m_c^3) I_1(3,3,1)+(8 m_s-16 m_c) I_2(1,2,3)
   \\ \nnb &+&
   (16 m_s-8 m_c) I_2(1,3,2)+(-16 m_c^3+16 m_s m_c^2-16 m_s^2 m_c+16 m_s^3)I_2(1,3,3)
   \\ \nnb &+&
   (32 m_b-16 m_c) I_2(2,3,1)+(16 m_b-32 m_c) I_2(3,2,1)+(32 m_b^3-32 m_c m_b^2
   \\ \nnb &+&
   32 m_c^2 m_b-32 m_c^3) I_2(3,3,1)\Bigg] +(8 m_c+8 m_s) I(1,1,3)-24 m_c m_s^2 I(1,1,4)
   \\ \nnb &+&
   8 m_s I(1,2,2)+(16 m_c^3+8 m_s m_c^2+8 m_s^2 m_c+8 m_s^3) I(1,2,3)
   \\ \nnb &+&
   (24 m_s-32 m_c) I(1,3,1)+(24 m_c^3+16 m_s^2 m_c) I(1,3,2)+(8 m_c^5-16 m_s^2 m_c^3
   \\ \nnb &+&
   8 m_s^4 m_c) I(1,3,3)-24 m_c^3 I(1,4,1)+(24 m_b-16 m_c+16 m_s) I(2,2,1)+(-16 m_c^3
   \\ \nnb &+&
   32 m_b m_c^2+40 m_s m_c^2-32 m_b^2 m_c+16 q^2 m_c-32m_b m_s m_c+48 m_b^2 m_s) I(2,3,1)
   \\ \nnb &+&
   (24 m_b-64 m_c+24 m_s) I(3,1,1)+(24 m_b^3-32 m_c m_b^2+40 m_s m_b^2+56 m_c^2 m_b
   \\ \nnb &+&
   8 q^2 m_b-32 m_c m_s m_b-32 m_c^3+16 m_c q^2+48 m_c^2 m_s) I(3,2,1)+(-16 m_c^5
   \\ \nnb &+&
   32 m_b m_c^4+24 m_s m_c^4-32 m_b^2 m_c^3+16 q^2 m_c^3-32 m_b m_s m_c^3+32 m_b^3 m_c^2+16 m_b^2 m_s m_c^2
   \\ \nnb &-&
   16m_b^4 m_c+16 m_b^2 q^2 m_c-32 m_b^3 m_s m_c+24 m_b^4 m_s) I(3,3,1)+72 m_b^2 m_c I(4,1,1)
   \\ \nnb &+&
   (-8 m_b+16 m_c+40 m_s) I_1(1,1,3)+(48 m_s^3+24 m_b m_s^2-72 m_c m_s^2) I_1(1,1,4)
   \\ \nnb &+&
   (16 m_s-16 m_c) I_1(1,2,2)+(32 m_c^3-16 m_s m_c^2+16 m_s^2 m_c-32 m_s^3) I_1(1,2,3)
   \\ \nnb &+&
   (-8 m_b-72 m_c-16 m_s) I_1(1,3,1)+(32 m_c^3-16 m_s m_c^2+16 m_s^2 m_c
   \\ \nnb &-&
   32 m_s^3) I_1(1,3,2)+(16 m_c^5-16 m_s m_c^4-32 m_s^2 m_c^3+32 m_s^3 m_c^2+16 m_s^4 m_c
   \\ \nnb &-&
   16 m_s^5) I_1(1,3,3)+(-72 m_c^3+24 m_b m_c^2+48 m_s m_c^2) I_1(1,4,1)
   \\ \nnb &+&
   (8 m_b-8 m_s) I_1(2,1,2)+(-56 m_s^3-24 m_b m_s^2-8 m_b^2 m_s+24 q^2 m_s
   \\ \nnb &+&
   16 m_b q^2) I_1(2,1,3)+(-16 m_b^3+8 m_c m_b^2-8 m_c^2 m_b-16 q^2 m_b+16 m_c^3
   \\ \nnb &+&
   8 m_c q^2) I_1(2,3,1)+(32 m_b-136 m_c+80 m_s) I_1(3,1,1)+(-8 m_b^3-16m_s m_b^2
   \\ \nnb &-&
   32 m_s^3+32 m_s q^2) I_1(3,1,2)+(-8 m_b^5-16 m_s m_b^4+16 m_s^2 m_b^3+16 q^2 m_b^3+32 m_s^3 m_b^2
   \\ \nnb &+&
   32 m_s q^2 m_b^2-8 m_s^4 m_b-8 q^4 m_b+16 m_s^2 q^2 m_b-16 m_s^5-16 m_s q^4+32 m_s^3 q^2) I_1(3,1,3)
   \\ \nnb &+&
   (-16 m_b^3+8 m_c m_b^2-8 m_c^2 m_b-8 q^2 m_b+16 m_c^3+16 m_c q^2) I_1(3,2,1)+(-8 m_b^5
   \\ \nnb &+&
   8 m_c m_b^4+16 m_c^2 m_b^3-16 q^2 m_b^3-16 m_c^3 m_b^2+16 m_c q^2 m_b^2-8 m_c^4 m_b-16 m_c^2 q^2 m_b
   \\ \nnb &+&
   8m_c^5+16 m_c^3 q^2) I_1(3,3,1)+(-72 m_b^3+216 m_c m_b^2-144 m_s m_b^2)I_1(4,1,1)
   \\ \nnb &+&
   (8 m_c+16 m_s) I_2(1,1,3)+(24 m_s^3-24 m_c m_s^2) I_2(1,1,4)+(8 m_s-8m_c) I_2(1,2,2)
   \\ \nnb &+&
   (16 m_c^3-8 m_s m_c^2+8 m_s^2 m_c-16 m_s^3) I_2(1,2,3)+(-24m_c-8 m_s) I_2(1,3,1)
   \\ \nnb &+&
   (16 m_c^3-8 m_s m_c^2+8 m_s^2 m_c-16 m_s^3)I_2(1,3,2)+(8 m_c^5-8 m_s m_c^4-16 m_s^2 m_c^3
   \\ \nnb &+&
   16 m_s^3 m_c^2+8 m_s^4 m_c-8m_s^5) I_2(1,3,3)+(24 m_c^2 m_s-24 m_c^3) I_2(1,4,1)
   \\ \nnb &+&
   (8 m_b-8 m_s)I_2(2,1,2)+(16 m_s q^2-40 m_s^3) I_2(2,1,3)+(8 m_b-48 m_c
   \\ \nnb &+&
   40 m_s) I_2(3,1,1)+(8m_b^3-8 m_s m_b^2+8 m_s^2 m_b-8 q^2 m_b-16 m_s^3+16 m_s q^2) I_2(3,1,2)
   \\ \nnb &+&
   (-8 m_s^5+16 m_b^2 m_s^3+16 q^2 m_s^3-8 m_b^4 m_s-8 q^4 m_s+16 m_b^2 q^2 m_s)I_2(3,1,3)
   \\ \nnb &+&
   (72 m_b^2 m_c-72 m_b^2 m_s) I_2(4,1,1)\Bigg\}
   \\ \nnb &+&
   D^1_0 \Bigg\{-24m_c^3 I(1,4,1) -24 m_s^2m_c I(1,1,4) +72 m_b^2 I(4,1,1) m_c+(24 m_b-16 m_c
   \\ \nnb &+&
   24 m_s) I(1,1,3)+(16 m_b-16 m_c+24 m_s) I(1,2,2)+(-32 m_c^3+48 m_b m_c^2+56 m_s m_c^2
   \\ \nnb &-&
   32 m_s^2 m_c+16 q^2 m_c-32 m_b m_s m_c+24 m_s^3+40 m_b m_s^2+8 m_sq^2) I(1,2,3)
   \\ \nnb &+&
   (24 m_b-32 m_c) I(1,3,1)+(-16 m_c^3+40 m_b m_c^2+32 m_s m_c^2-32m_s^2 m_c+16 q^2 m_c
   \\ \nnb &-&
   32 m_b m_s m_c+48 m_b m_s^2) I(1,3,2)+(-16 m_c^5+24 m_b m_c^4+32 m_s m_c^4-32 m_s^2 m_c^3
   \\ \nnb &+&
   16 q^2 m_c^3-32 m_b m_s m_c^3+32 m_s^3 m_c^2+16 m_b m_s^2 m_c^2-16 m_s^4 m_c-32 m_b m_s^3 m_c
   \\ \nnb &+&
   16 m_s^2 q^2 m_c+24 m_bm_s^4) I(1,3,3)+(8 m_b+8 m_c) I(2,2,1)
   \\ \nnb &+&
   (24 m_c^3+16 m_b^2 m_c) I(2,3,1)+(8m_b-40 m_c) I(3,1,1)+(8 m_b^3+8 m_c m_b^2+8 m_c^2 m_b
   \\ \nnb &+&
   16 m_c^3)I(3,2,1)+(8 m_c^5-16 m_b^2 m_c^3+8 m_b^4 m_c) I(3,3,1)+(8 m_s-8 m_b)I_1(1,1,3)
   \\ \nnb &+&
   (24 m_b m_s^2-24 m_c m_s^2) I_1(1,1,4)+(-8 m_b-24 m_c)I_1(1,3,1)+(24 m_b m_c^2
   \\ \nnb &-&
   24 m_c^3) I_1(1,4,1)+(8 m_s-8 m_b) I_1(2,1,2)+(24m_s^3-24 m_b m_s^2-8 m_b^2 m_s-8 q^2 m_s
   \\ \nnb &+&
   16 m_b q^2) I_1(2,1,3)+(-16 m_b^3+8 m_cm_b^2-8 m_c^2 m_b+16 m_c^3) I_1(2,3,1)
   \\ \nnb &+&
   (16 m_b-40 m_c) I_1(3,1,1)+(-24 m_b^3-16m_s^2 m_b+16 q^2 m_b) I_1(3,1,2)
   \\ \nnb &+&
   (-8 m_b^5+16 m_s^2 m_b^3+16 q^2 m_b^3-8 m_s^4m_b-8 q^4 m_b+16 m_s^2 q^2 m_b) I_1(3,1,3)
   \\ \nnb &+&
   (-16 m_b^3+8 m_c m_b^2-8 m_c^2 m_b+16m_c^3) I_1(3,2,1)+(-8 m_b^5+8 m_c m_b^4+16 m_c^2 m_b^3
   \\ \nnb &-&
   16 m_c^3 m_b^2-8 m_c^4m_b+8 m_c^5) I_1(3,3,1)+(72 m_b^2 m_c-72 m_b^3) I_1(4,1,1)+(-16 m_b
   \\ \nnb &+&
   8 m_c+32 m_s) I_2(1,1,3)+(24 m_s^3+48 m_b m_s^2-72 m_c m_s^2) I_2(1,1,4)
   \\ \nnb &+&
   (8 m_s-8 m_c)I_2(1,2,2)+(16 m_c^3-8 m_s m_c^2+8 m_s^2 m_c+16 q^2 m_c-16 m_s^3
   \\ \nnb &-&
   8 m_s q^2)I_2(1,2,3)+(-16 m_b-72 m_c-8 m_s) I_2(1,3,1)+(16 m_c^3-8 m_s m_c^2+8 m_s^2 m_c
   \\ \nnb &+&
   8 q^2 m_c-16 m_s^3-16 m_s q^2) I_2(1,3,2)+(8 m_c^5-8 m_s m_c^4-16 m_s^2 m_c^3+16 q^2m_c^3
   \\ \nnb &+&
   16 m_s^3 m_c^2-16 m_s q^2 m_c^2+8 m_s^4 m_c+16 m_s^2 q^2 m_c-8 m_s^5-16 m_s^3q^2) I_2(1,3,3)
   \\ \nnb &+&
   (-72 m_c^3+48 m_b m_c^2+24 m_s m_c^2) I_2(1,4,1)+(8 m_s-8 m_b)I_2(2,1,2)+(8 m_s^3
   \\ \nnb &-&
   48 m_b m_s^2-16 m_b^2 m_s+32 m_b q^2) I_2(2,1,3)+(-32m_b^3+16 m_c m_b^2-16 m_c^2 m_b
   \\ \nnb &+&
   32 m_c^3) I_2(2,3,1)+(40 m_b-128 m_c+40 m_s)I_2(3,1,1)+(-40 m_b^3-8 m_s m_b^2
   \\ \nnb &-&
   24 m_s^2 m_b+24 q^2 m_b-16 m_s^3+16 m_s q^2)I_2(3,1,2)+(-16 m_b^5-8 m_s m_b^4+32 m_s^2 m_b^3
   \\ \nnb &+&
   32 q^2 m_b^3+16 m_s^3 m_b^2+16 m_sq^2 m_b^2-16 m_s^4 m_b-16 q^4 m_b+32 m_s^2 q^2 m_b-8 m_s^5
   \\ \nnb &-&
   8 m_s q^4+16 m_s^3 q^2)I_2(3,1,3)+(-32 m_b^3+16 m_c m_b^2-16 m_c^2 m_b+32 m_c^3) I_2(3,2,1)
   \\ \nnb &+&
   (-16m_b^5+16 m_c m_b^4+32 m_c^2 m_b^3-32 m_c^3 m_b^2-16 m_c^4 m_b+16 m_c^5)I_2(3,3,1)
   \\  &+&(-144 m_b^3+216 m_c m_b^2-72 m_s m_b^2) I_2(4,1,1)\Bigg\}\eea

where
 \bea D_i^j \left[I_n(M_1^2,M_2^2)\right]&=& ( M_1^2)^i (M_2^2)^j \frac{\partial_i}{\partial( M_1^2)^i} \frac{\partial^j}
{\partial( M_2^2 )^j}\left[( M_1^2)^i (M_2^2 )^j I_n(M_1^2,M_2^2) \right]~.
\nnb \eea

\begin{thebibliography}{9}

\bibitem{Cogollo:2013mga}
  D.~Cogollo, F.~S.~Queiroz and P.~Vasconcelos,
  arXiv:1312.0304 [hep-ph].
%
\bibitem{Li:2013vlx}
  X.~-Q.~Li, Y.~-D.~Yang and X.~-B.~Yuan,
  arXiv:1311.2786 [hep-ph].
\bibitem{Buras:2013qja}
  A.~J.~Buras and J.~Girrbach,
  FLAVOUR(267104)-ERC-50
  [arXiv:1309.2466 [hep-ph]].
\bibitem{Ali:2013zfa}
  A.~Ali, A.~Y.~.Parkhomenko and A.~V.~Rusov,
  arXiv:1312.2523 [hep-ph].

\bibitem{Richard:2013xfa}
  F.~ço.~Richard,
  arXiv:1312.2467 [hep-ph].
\bibitem{Branco}
  G.~C.~Branco and M.~N.~Rebelo,
  PoS Corfu {\bf 2012}, 024 (2013)
  [arXiv:1308.4639 [hep-ph]].

\bibitem{Aaij:2013qta}
  RAaij {\it et al.}  [LHCb Collaboration],
  Phys.\ Rev.\ Lett.\  {\bf 111}, 191801 (2013)
  [arXiv:1308.1707 [hep-ex]].

\bibitem{Aaij:2013dgw}
  R.~Aaij {\it et al.}  [LHCb Collaboration],
  Phys.\ Rev.\ Lett.\  {\bf 111}, 151801 (2013)
  [arXiv:1308.1340 [hep-ex]].

\bibitem{Aaij:2013aka}
  RAaij {\it et al.}  [LHCb Collaboration],
  Phys.\ Rev.\ Lett.\  {\bf 111}, 101805 (2013)
  [arXiv:1307.5024 [hep-ex]].
\bibitem{Vesterinen:2013jia}
  M.~Vesterinen [LHCb Collaboration],
  PoS Beauty {\bf 2013}, 005 (2013)
  [arXiv:1306.0092 [hep-ex]].

\bibitem{Aaij:2013iag}
  R.~Aaij {\it et al.}  [LHCb Collaboration],
  JHEP {\bf 1308}, 131 (2013)
  [arXiv:1304.6325 [hep-ex]].
\bibitem{Bashiry:2013waa}
  V.~Bashiry,
  arXiv:1305.6535 [hep-ph].
\bibitem{Ghahramany:2013gya}
  N.~Ghahramany and A.~R.~Houshyar,
  Acta Phys.\ Polon.\ B {\bf 44}, no. 9, 1857 (2013).

\bibitem{Gan:2012tt}
  L.~-F.~Gan, Y.~-L.~Liu, W.~-B.~Chen and M.~-Q.~Huang,
  Commun.\ Theor.\ Phys.\  {\bf 58}, 872 (2012)
  [arXiv:1212.4671 [hep-ph]].

\bibitem{Sarac:2013rpa}
  Y.~Sarac, K.~Azizi and H.~Sundu,
  Nucl.\ Phys.\ Proc.\ Suppl.\  {\bf 245}, 164 (2013).

\bibitem{Khodjamirian:2009ys}
  A.~Khodjamirian, C.~.Klein, T.~.Mannel and N.~Offen,
  Phys.\ Rev.\ D {\bf 80}, 114005 (2009)
  [arXiv:0907.2842 [hep-ph]].

\bibitem{Aliev:2006vs}
  T.~M.~Aliev and M.~Savci,
  Eur.\ Phys.\ J.\ C {\bf 47}, 413 (2006)
  [hep-ph/0601267].

\bibitem{Azizi:2008vv}
  K.~Azizi, F.~Falahati, V.~Bashiry and S.~M.~Zebarjad,
  Phys.\ Rev.\ D {\bf 77}, 114024 (2008)
  [arXiv:0806.0583 [hep-ph]].

  \bibitem{Marques de Carvalho:1999ia}
  R.~S.~Marques de Carvalho, F.~S.~Navarra, M.~Nielsen, E.~Ferreira and H.~G.~Dosch,
  Phys.\ Rev.\ D {\bf 60}, 034009 (1999)
  [hep-ph/9903326].




\bibitem{Buras:1994dj}
  A.~J.~Buras and M.~Munz,
  Phys.\ Rev.\  D {\bf 52}, 186 (1995)
  [arXiv:hep-ph/9501281].

\bibitem{Lim:1988yu}
  C.~S.~Lim, T.~Morozumi and A.~I.~Sanda,
  Phys.\ Lett.\  B {\bf 218}, 343 (1989).



\bibitem{Willey} W. S. Hou, R. S. Willey and A. Soni, {\it Phys. Rev. Lett} {\bf
58} (1987) 1608; {\it ibid} {\bf 60} (1988) 2337 {\it Erratum}.

\bibitem{deshpande}  N. G. Deshpande and J. Trampetic, {\it Phys. Rev. Lett} {\bf
60} (1988) 2583.
\bibitem{R5734} M. Jezabek and J. H. K\"{u}hn,
{\it Nucl. Phys.} {\bf B320} (1989) 20.
\bibitem{misiak} M.\,Misiak, Nucl. Phys. {\bf B393}\, {1993} 23.
\bibitem{misiakE}M.\,Misiak, Nucl. Phys. {\bf B439}\, 461(E) \,{1995}.
\bibitem{NNLL}T. Huber, E. Lunghi, M. Misiak and D. Wyler, {\it Nucl. Phys. }{\bf
B740} (2006) 105.

\bibitem{pdg12}J. Beringer et al. (Particle Data Group), Phys. Rev. D86, 010001 (2012)
\bibitem{Ali:1999mm}
  A.~Ali, P.~Ball, L.~T.~Handoko and G.~Hiller,
  Phys.\ Rev.\  D {\bf 61}, 074024 (2000)
  [arXiv:hep-ph/9910221].

\bibitem{Ali:1991is}
  A.~Ali, T.~Mannel and T.~Morozumi,
  Phys.\ Lett.\  B {\bf 273}, 505 (1991).


\bibitem{Kruger:1996cv}
  F.~Kruger and L.~M.~Sehgal,
  Phys.\ Lett.\  B {\bf 380}, 199 (1996)
  [arXiv:hep-ph/9603237].

  \bibitem{R7320} V. A. Fock, Sov. Phys. {\bf 12}, 404 (1937).
\bibitem{R7321} J. Schwinger, Phys. Rev. {\bf 82}, 664 (1951).
\bibitem{R7322} V. Smilga, Sov. J. Nucl. Phys. {\bf 35}, 215 (1982).
\bibitem{R7323} V. V. Kiselev, A. K. Likhoded,  A. I. Onishchenko,
 Nucl. Phys.  {\bf B 569}  (2000) 473.

\bibitem{Shifman1} M. A. Shifman, A. I. Vainshtein,  V. I. Zakharov, Nucl.
Phys. {\bf B147}  (1979) 385.

\bibitem{Bashiry:2011pp}
  V.~Bashiry, K.~Azizi and S.~Sultansoy,
  Phys.\ Rev.\ D {\bf 84}, 036006 (2011)
  [arXiv:1104.2879 [hep-ph]].

\bibitem{Wang:2012kw}
  Z.~-G.~Wang,
  Eur.\ Phys.\ J.\ A {\bf 49} (2013) 131
  [arXiv:1203.6252 [hep-ph]].

\bibitem{AbdElHady:1998kc}
  A.~Abd El-Hady, M.~A.~K.~Lodhi and J.~P.~Vary,
  Phys.\ Rev.\ D {\bf 59}, 094001 (1999)
  [hep-ph/9807225].
\bibitem{Kiselev:2000jc}
  V.~V.~Kiselev, A.~E.~Kovalsky and A.~I.~Onishchenko,
  Phys.\ Rev.\ D {\bf 64}, 054009 (2001)
  [hep-ph/0005020].
\end{thebibliography}
\end{document}